\documentstyle[12pt]{article}
\date{}
\def\con{{}_{\_\rule{-1pt}{0pt}\_}
\rule{-2pt}{0pt}\raise1.5pt\hbox{$\mid$}\hspace{2pt}}

\newcommand{\beq}{\begin{eqnarray}}
\newcommand{\eeq}{\end{eqnarray}}
\newtheorem{DF}{DEFINITION}
\newtheorem{TH}{THEOREM}
\newtheorem{PR}{PROPOSITION}
\newcommand{\om}{\mbox{\boldmath $\omega$}}

\title{ Hamiltonian Structure for Classical
       Electrodynamics of a Point Particle}
\author{ Dariusz Chru\'sci\'nski\footnotemark \\
        {\it Fakult\"{a}t f\"{u}r Physik, Universit\"at Freiburg}\\
        {\it Hermann-Herder Str. 3,
        D-79104 Freiburg, Germany}}

\begin{document}

\def\thefootnote{\relax}\footnotetext{$^*$Alexander von Humboldt fellow.
Permanent address:
       Institute of Physics,
        Nicholas Copernicus University,
        ul. Grudzi\c{a}dzka 5/7, 87-100 Toru\'n, Poland.}

\begin{sloppypar}
\maketitle

\begin{abstract}

We prove that, contrary to the common belief, the classical Maxwell
electrodynamics of a point-like particle may be formulated
 as an infinite-dimensional
Hamiltonian system. We derive well defined quasi-local Hamiltonian
which possesses direct physical interpretation being equal to the total energy
of the composed system (field + particle). The phase space of this system is
endowed with an interesting symplectic structure. We prove that this
structure is strongly non-degenerated and, therefore, enables one to define
the consistent Poisson bracket for particle's and field degrees of freedom. We
strees that this formulation is perfectly gauge-invariant.
\end{abstract}

\newpage

\def\theequation{\thesection.\arabic{equation}}

\section{Introduction}
\setcounter{equation}{0}

Classical electrodynamics of charged, point-like particles  is
usually based on the celebrated Lorentz-Dirac equation 
\cite{Dirac1} which, although very
useful in many applications, has certain inherent difficulties.
  There have been many attempts to
solve this problem (see e.~g. \cite{Rohrlich} - \cite{Parrot}
for the rewiev). However, there is no solution which is accepted by all
physicists.

The aim of this paper is to show that despite of these problems the classical
electrodynamics of point particles have well defined Hamiltonian structure. We
would like to stress that the
existence of this structure is nontrivial (up to our
knowledge, it is completely unknown).

The derivation of  the Lorentz-Dirac equation is based on
the decomposition of the
electromagnetic field into ``retarded'' and ``incoming'' components:
\beq
f^{\mu\nu} = f_{ret}^{\mu\nu} + f_{in}^{\mu\nu}\ .
\eeq
In this approach it is impossible to correctly formulate the initial value
problem and  find the generator of time translations (see \cite{Rohrlich}).
The dynamics of external ``incoming'' field is already given and the retarded
component is fully determined by the particle's motion. Therefore, the field
degrees of freedom are completely eliminated and there is simply no room for
the Hamiltonian description.
However, recently it was shown (cf. \cite{KIJ})  that classical electrodynamics
of point-like objects may be formulated as an infinite-dimensional dynamical
system. In this approach particles and field degrees of freedom are kept on the
same footing. There is no equations of motion for the particle. It was shown
that the conservation law for the total (renormalized) four-momentum is
equivalent to a certain boundary condition for the electromagnetic field along
the particle's trajectory.
 Together with this
condition, the theory (called {\it electrodynamics of
moving particles}) becomes causal and complete: initial
data for both the field and the particles uniquely imply
the evolution of the system. This means e.~g.~that the
particles trajectories may also be calculated uniquely from
the initial data.
Because this approach is relatively little known
(and it is crucial for the present paper) we present a short review in the next
Section.

It turns out (cf. \cite{Kij-Dar}) that the
electrodynamics of moving particles possesses very
interesting Lagrangian structure.  Let us note that the
standard variational principle used in electrodynamics
cannot be extended to the theory containing also point-like
particles interacting with the electromagnetic field. Such
a principle is based on the following Lagrangian, written
usually in textbooks:
\begin{eqnarray}   \label{ill}
L_{total} = L_{Maxwell}+ L_{particle} + L_{int} \ ,
\end{eqnarray}
with
\begin{eqnarray}
L_{Maxwell} = -\frac{1}{4}\sqrt{-g}f^{\mu \nu}f_{\mu \nu}\,
,\label{ML}
\end{eqnarray}
\begin{eqnarray}
L_{particle} := - m  \delta_\zeta \ , \label{Lparticle}
\end{eqnarray}
and the interaction term given by
\begin{eqnarray}
L_{int} := e A_\mu u^\mu \delta_\zeta \ .
\label{Linteraction}
\end{eqnarray}
Here by $\delta_\zeta$ we denote the Dirac delta
distribution localized on the particle trajectory $\zeta$.
The above Lagrangian may be used to derive the trajectories
of the test particles, when the field is given {\it a
priori}. In a different context, it may also be used to
derive Maxwell equations, if the particle trajectories are
given {\it a priori}. Simultaneous variation with respect
to both fields and particles leads, however, to a
contradiction, since the Lorentz force will be always ill
defined due to Maxwell equations.

But already in the context of the inhomogeneous Maxwell
theory with given point-like sources, the variational
principle based on Lagrangian (\ref{ill}) is of very
limited use, since the interaction term $L_{int}$ becomes
infinite. As a consequence, the Hamiltonian of such a
theory will always be ill defined, although the theory
displays a perfectly causal behaviour.

To improve this bad feature of the theory, a new, quasi-local
variational principle for the Maxwell field with given sources was proposed
in \cite{Kij-Dar}.   This new
variational principle is based on the quasi-local
Lagrangian which, contrary to (\ref{ill}), is well defined,
i.e.  produces no infinities. It was proved (see
\cite{Kij-Dar}) that adding to that Lagrangian the particle
Lagrangian (\ref{Lparticle}) and varying it with respect to
both fields and particles is now possible and does not lead
to any contradiction.  As a result, one obtains precisely
the electrodynamics of moving particles proposed in
\cite{KIJ}.

In the present paper we give the Hamiltonian formulation of
the above theory, i.~e. we shall prove that electrodynamics
of moving particles may be formulated as an infinite-dimensional
Hamiltonian system. For the simplicity we consider here only one particle case.
It is of course possible to generalise this result to many particles, however,
it is technically much more complicated.

 Obviously, the Hamiltonian description
based on the standard Lagrangian (\ref{ill}) is
inconsistent in the case of a point-like sources.
In particular, the Hamiltonian of such a theory is always
ill defined.
Our approach, based on the quasi-local Lagrangian
 defined in \cite{Kij-Dar} leads {\it via} an
appropriate Legendre transformation to the well defined
quasi-local Hamiltonian structure  for fields interacting
with a charged particle. In particular, the Hamiltonian of the composed
``particle + field'' system  equals
numerically to the total energy of the composed system and,
therefore, has a direct physical interpretation.

Moreover, it turns out that the above theory possesses
highly nontrivial Poisson bracket structure. This structure
is defined {\it via} the symplectic form living in the
phase space of the entire system (particle + field).  The
characteristic feature of the infinite-dimensional
symplectic manifold is that the symplectic 2-form is in
general only  {\it weakly nondegenerate}, cf.
\cite{Marsden}. It means that there need not exist a Hamiltonian vector field
$X_{{\cal F}}$ corresponding to every given functional $\cal F$ on the phase
space of the system. For example, our system has well defined Hamiltonian,
however, corresponding vector field generated dynamics is not defined
throughout the phase space.
 We prove (cf. Theorem~\ref{TH1}) that this vector field
 is well defined if and only if we restrict the phase space to the states
 fulfilling
 the fundamental equation for electrodynamics of moving
 particles.

Our main result consists in the Theorem ~\ref{strong} which says, that on the
reduced phase space the symplectic 2-form
 becomes
{\it strongly nondegenerate}. This nice mathematical result enables one to
define the  Poisson bracket structure in the space of
functionals over the reduced phase space.
To our knowledge the above Poisson bracket structure is the
first fully consistent structure for the theory of
interacting particles and fields, i.~e.  when the particles
and fields variables are kept at the same footing.

The paper is organized as follows.

Section~\ref{Statement} contains the main results of the
new approach to classical electrodynamics of point particles proposed in
\cite{KIJ}.

In Section~\ref{Comoving} we present a new technique,
developed in
\cite{Kij-Dar},
which enables us to describe at the same footing the field
and the particle's degrees of freedom.  For this purpose we
formulate the Hamiltonian structure of any relativistic,
hyperbolic field theory with respect to a non-inertial
reference frame defined as a rest-frame for an arbitrarily
moving observer.

In Section~\ref{Renorm} we show how to extend the above
approach to the case of electrodynamical field interacting
with point particles.

For the reader convenience we present in
Section~\ref{Fluxes} a new Hamiltonian structure for
Maxwell electrodynamics. The new electrodynamical
Hamiltonian corresponds to a symmetric energy-momentum
tensor and  it is related {\it via} a simple Legendre
transformation with electrodynamical Lagrangian derived in
\cite{Kij-Dar}.

Next three Sections present consistent Hamiltonian structure
for electrodynamics of a point particle, i.~e. we
derive the quasi-local Hamiltonian and find the well
defined formula for the Poisson bracket. In
Section~\ref{Poincare} we show that the above Poisson
bracket structure is consistent with Poincar\'e algebra
structure of relativistic theory.

Finally, in Section~\ref{External} we present the
Hamiltonian formulation for the particle interacting not
only with the radiation field, but also with a fixed,
external potential, produced by a heavy external device.
This is a straightforward extension of our theory, where
the homogeneous boundary condition for the radiation field
is replaced by an inhomogeneous condition, the
inhomogeneity being provided by the external field.

\section{The new approach to electrodynamics of a point particle}
 \label{Statement}
\setcounter{equation}{0}

In the present Section we briefly sketch the new approach
to electrodynamics of a point particle presented in
\cite{KIJ}.

Let ${\bf y} = {\bf q}(t)$ with $t = y^0$, be the
coordinate description of a time-like world line $\zeta$ of
a charged particle with respect to a laboratory frame, i.e.
to a system $(y^{\mu}),\
\mu = 0,1,2,3;$ of Lorentzian space-time coordinates.

The theory contains as a main part the standard Maxwell
equations with point-like sources:
\begin{eqnarray}  \label{Maxeq}
\partial_{[\lambda}f_{\mu\nu]} &=& 0\ ,\nonumber\\
\partial_{\mu}f^{\mu\nu} &=& eu^{\nu}\delta_{\zeta} \ ,
\end{eqnarray}
where $u^{\nu}$ stands for the particle four-velocity and
$\delta_{\zeta}$ denotes the $\delta$-distribution
concentrated on the smooth world line $\zeta$:
\begin{eqnarray}
\delta_{\zeta}(y^0,y^k) = \sqrt{1 - ({\bf v}(y^0))^2}\, \delta^{(3)}
(y^k - q^k(y^0)) \ .\nonumber
\end{eqnarray}
Here ${\bf v} = (v^k) = (\dot{q}^k)$ is the corresponding
3-velocity and ${\bf v}^2$ denotes the square of its
3-dimensional length (we use the Heaviside-Lorentz system
of units with $c = 1$).  In the case of many particles the
total current is a sum of contributions corresponding to
many disjoint world lines and the value of charge is
assigned to each world line separately.

For a given particle's trajectory, equations (\ref{Maxeq})
define a deterministic theory. This means that initial data
for the electromagnetic field uniquely determine its
evolution. However, if we want to treat also the particle`s
initial data ({\bf q},{\bf v}) as dynamical variables, the
theory based on the Maxwell equations alone is no longer
deterministic: the particle`s trajectory can be arbitrarily
modified in the future or in the past without changing the
initial data.

This non-completeness of the theory is usually attributed
to the fact that the particle's equations of motion are
still missing. However, it was proved in
\cite{KIJ} that the field initial data fully determine the particle's
acceleration and this is due to Maxwell equations only,
without postulating any equations of motion. More
precisely, there is a one-to-one correspondence between the
($r^{-1}$)-term of the field in the vicinity of the
particle and the acceleration of the particle.  The easiest
way to describe this property of Maxwell theory is to use
the particle's rest-frame.  For this purpose consider the
3-dimensional hyperplane  $\Sigma_t$ orthogonal to $\zeta$
at the point $(t,{\bf q}(t))\in \zeta$.  We shall call
$\Sigma_t$ the ``rest frame hyperplane''.  Choose on
$\Sigma_t$ any system $(x^i)$ of cartesian coordinates
centered at the particle's position and denote by $r$ the
corresponding radial coordinate.  The initial data for the
field on $\Sigma_t$ are given by the electric induction
field ${\bf D}=(D^i)$ and the magnetic induction field
${\bf B}=(B^i)$ fulfilling the conditions ${\bf div}\: {\bf
B} = 0$  and  ${\bf div}\: {\bf D} = e\: {\delta}^{(3)}_0$.
Maxwell equations can be solved for arbitrary data,
fulfilling the above constraints, but the solution will be
usually non-regular, even far away from the particles. To
avoid singularities propagating over a light cone from
$(t,{\bf q}(t))$, the singular part of the data in the
vicinity of the particle has to be equal to
\begin{eqnarray}
D^k=\frac e{4\pi} \left[
\frac {x^k}{r^3} - \frac 1{2r} \left(
a_i \frac {x^i x^k}{r^2} + a^k \right) \right] + O(1) \ ,
\label{regular}
\end{eqnarray}
where ${\bf a} = (a^k)$ is the acceleration of the particle
(in the rest frame we have $a^0 = 0$) and $O(1)$ denotes
the nonsingular part of the field (the magnetic field $B^k
(r)$ cannot contain any singular part). This gives the
one-to-one correspondence between the ($r^{-1}$)-term of
the field and the particle's acceleration, which is implied
by the regularity of the field outside of the trajectory
$\zeta$.

Hence, for regular solutions, the time derivatives
$(\dot{\bf D},\dot{\bf B},\dot{\bf q},\dot{\bf v})$ of the
Cauchy data $({\bf D},{\bf B},{\bf q},{\bf v})$ of the
composed (fields + particle) system are uniquely
determined by the data themselves. Indeed, $\dot{\bf D}$
and $\dot{\bf B}$ are given by the Maxwell equations,
$\dot{\bf q} = {\bf v}$ and $\dot{\bf v}$ may be uniquely
calculated from equation (\ref{regular}).  Nevertheless,
the theory is not complete and its evolution is not
determined by the initial data.  This non-completeness may
be interpreted as follows.  Field evolution takes place not
in the entire Minkowski space $M$, but only outside the
particle, i.~e.~in a manifold with a non-trivial boundary
$M_\zeta := M - \{\zeta\} $. The boundary conditions for
the field are still missing.

To find this missing condition, the following method was
used.  Guided by an extended particle model, an ``already
renormalized'' formula was proposed in \cite{KIJ}, which
assigns to each point $(t,q^k(t))$ of the trajectory a
four-vector $p^\lambda (t)$ according to
\begin{DF}
\beq   \label{4-momentum}
p_\lambda(t) := mu_\lambda(t) + \mbox{P}\int_{\Sigma}
 (T^\mu_{\ \lambda}
 - {{\bf T}_{(t)}}^\mu_{\ \lambda})\,d\Sigma_\mu\ ,
 \eeq
where $T^\mu_{\ \nu}$ denotes the symmetric energy-momentum tensor of the
Maxwell field and ${{\bf T}_{(t)}}^\mu_{\ \nu}$ denotes the energy-momentum
tensor corresponding to the electromagnetic field produced by the
uniformly moving particle along the stright line
tangent to the trajectory $\zeta$ at $(t,{\bf q}(t))$. ``P'' denotes the
principal value of the singular integral and $\Sigma$ is an arbitrary
space-like hypersurface passing through the point $(t,{\bf q}(t))$.
\end{DF}

 It was proved in \cite{KIJ} that the four-vector (\ref{4-momentum})
 is well defined and does not depend on the particular choice of $\Sigma$.
 It is interpreted as the total
four-momentum of the physical system composed of both the
particle and the field.  For a generic trajectory $\zeta$ and a generic
solution of Maxwell equations (\ref{Maxeq}) this quantity
is not conserved, i.~e.~it depends upon $t$.  The
conservation law
\begin{eqnarray}   \label{conservation4}
\frac{d}{dt}\, p^\lambda (t) = 0
\end{eqnarray}
is proposed as an additional equation, which completes the
theory. It was shown that, due to Maxwell equations, only 3
among the 4 equations (\ref{conservation4}) are
independent. Given a laboratory reference frame, one may
take e.~g.~the conservation of the momentum ${\bf p} =
(p^k)$:
\begin{eqnarray}   \label{conservation}
\frac{d}{dt}\, {\bf p} = 0\
\end{eqnarray}
as independent equations. They already imply the energy
conservation
$\frac{d}{dt}\, p^0 (t) = 0 $.
In Section \ref{Reduction} we will show, that the above
momentum ${\bf p}$ is equal to
 the  momentum canonically conjugate to
the position of the particle, whereas $p_0$ is equal to the
total Hamiltonian of the composed (particle + field)
system.

 It has
been proved in \cite{KIJ} that, due to Maxwell equations,
the integral (global) condition (\ref{conservation}) is
equivalent to a (local) boundary condition for the
behaviour of the Maxwell field in the vicinity of the
trajectory. The condition was called {\it the fundamental
equation} of the electrodynamics of moving particles. In
particle's reference frame it may be formulated as a
relation between the $(r^{-1})$ and the $(r^{0})$ terms in
the expansion of the radial component of the electric field
in the vicinity of the particle:
\begin{eqnarray}
D^r(r)=\frac 1{4\pi }\left( \frac e{r^2} + \frac {\alpha}r
\right) +  \beta  + O(r) \ , \label{expansion}
\end{eqnarray}
where by $O(r)$ we denote terms vanishing for $r
\rightarrow 0$ like $r$ or faster. For a given value of $r$
both sides of (\ref{expansion}) are functions of the angles
(only the $r^{-2}$ term is angle--independent). The
relation between the acceleration and the $(r^{-1})$ --
term of the electric field given in (\ref{regular}) may be
rewritten in terms of the component $\alpha$ of this
expansion:
\begin{eqnarray}   \label{Dip}
\alpha = - e a_i \frac{x^i}{r}
\end{eqnarray}
(it implies that the quadrupole and the higher harmonics of
$\alpha$ must vanish for regular solutions).   One can prove
the following theorem (see \cite{KIJ}):
\begin{TH}
The conservation law
(\ref{conservation}) is equivalent to the following boundary condition
\begin{eqnarray}   \label{fund}
\mbox{DP}(m\alpha + e^2\beta) = 0\ ,
\end{eqnarray}
where DP($f$) denotes the dipole part of the function
$f$ on the sphere $S^2$.
\end{TH}
The main result of \cite{KIJ} may be summarised in the following
\begin{TH}
Maxwell equations together with the boundary condition (\ref{fund})
define complete, causal and fully
deterministic theory: initial data for particles and fields
uniquely determine the entire history of the system.
\end{TH}

\section{Hamiltonian  description of a relativistic field theory in the
co-moving frame}
\label{Comoving}
\setcounter{equation}{0}

\subsection{Definition of the co-moving frame}
To construct the Hamiltonian formulation of the above
theory, we will need a Hamiltonian description of
electrodynamics with respect to the particle's rest-frame.
In \cite{Kij-Dar} it was shown how to extend the standard
variational formulation of field theory to the case of
non-inertial frames (for the definition of the co-moving frame see also
\cite{MTW}).  In the present Section we show how to
extend the standard Hamiltonian formulation of a classical
field theory.

Consider any field theory based on a first-order
relativistically-invariant Lagrangian density
\begin{eqnarray}
L = L(\psi , \partial \psi) \ ,
\end{eqnarray}
where $\psi$ is a (possibly multi-index) field variable,
which we do not need to specify at the moment. As an
example, $\psi$ could denote a scalar, a spinor or a tensor
field.

We will describe the above field theory with respect to
accelerated reference frames, related with observers moving
along arbitrary space-time trajectories. Let $\zeta$ be
such a (time-like) trajectory, describing the motion of our
observer. Let  $y^k = q^k(t)$,
$k=1,2,3$; be the description of $\zeta$ with respect to a
laboratory reference frame, i.~e.~to a system
$(y^\lambda)$, $\lambda = 0,1,2,3$; of Lorentzian
space-time coordinates.
We will construct an accelerated reference frame, co-moving
with $\zeta$. For this purpose let us consider at each
point $(t,{\bf q}(t))\in \zeta$ the 3-dimensional
hyperplane  $\Sigma_t$ orthogonal to $\zeta$, i.e.
orthogonal to the four-velocity vector $U(t)=(u^{\mu}(t))$:
\begin{eqnarray}
(u^\mu) = (u^0 , u^k) := \gamma
 (1,v^k)  \ ,
\end{eqnarray}
where $v^k := {\dot q}^k$ and the relativistic factor
$\gamma := 1/ \sqrt {1-{\bf v}^2}$.  We shall call $\Sigma_t$ the
``rest frame surface''. Choose on $\Sigma_t$ any system
$(x^i)$ of {\it cartesian} coordinates, such that the
particle is located at its origin (i.~e.~at the point
$x^i=0$).

Let us consider space-time as a disjoint sum of rest frame
surfaces $\Sigma_t$, each of them corresponding to a
specific value of the coordinate $x^0 := t$ and
parameterized by the coordinates $(x^i)$. This way we obtain
a system $(x^\alpha)=(x^0,x^k)$ of ``co-moving''
coordinates in a neighbourhood of $\zeta$. Unfortunately,
it is not always a global system because different
$\Sigma$'s may intersect.  Nevertheless, we will use it
{\it globally} to describe the evolution of the field
$\psi$ from one $\Sigma_t$ to another. For a hyperbolic
field theory, initial data on one $\Sigma_t$ imply the
entire field evolution. We are allowed, therefore, to
describe this evolution as a 1-parameter family of field
initial data over subsequent $\Sigma$'s.

Formally, we will proceed as follows. We consider an
abstract space-time ${\bf M}:=T \times \Sigma $ defined as
the product of an abstract time axis $T={\bf R}^1$ with an
abstract, three dimensional Euclidean space $\Sigma = {\bf
R}^3$. Given a world-line $\zeta$, we will need an
identification of points of {\bf M} with points of physical
space-time $M$. Such an identification is not unique
because on each $\Sigma_t$ we have still the freedom of an
O(3)-rotation.

Suppose, therefore, that an identification $F$ has been
chosen, which is {\it local} with respect to the observer's
trajectory.  By locality we mean that, given the position
and the velocity of the observer at the time $t$, the
isometry
\begin{eqnarray}
F_{({\bf q}(t),{\bf v}(t))} : {\Sigma} \mapsto \Sigma_t
\label{isometry}
\end{eqnarray}
is already defined, which maps $0 \in \Sigma$ into the
particle position $(t,{\bf q}(t))\in \Sigma_t$.

As an example of such an isometry which is {\it local} with
respect to the trajectory we could take the one obtained as
follows.  Choose the unique boost transformation relating
the laboratory time axis $\partial/\partial y^0$ with the
observer's proper time axis $U$.  Next, define the position
of the $\partial/\partial x^k$ - axis on $\Sigma_t$ by
transforming the corresponding $\partial/\partial y^k$ --
axis of the laboratory frame by the same boost.  It is easy
to check, that the resulting formula for $F$ reads:
\begin{eqnarray}
y^0(t,x^l) & := & t +  \gamma(t)
x^l v_l(t)
\ ,\nonumber
\\ y^k(t,x^l) & := & q^k(t) +
\left( {\delta}^k_l + \varphi ({\bf v}^2)v^k v_l \right) x^l
 \ . \label{embedding}
\end{eqnarray}
Here, the following function of a real variable has been
used:
\begin{eqnarray}
\varphi (\tau):= \frac 1{\tau}
\left( \frac {1}{\sqrt{1 - {\tau}}} \ - 1 \right) =
\frac 1{\sqrt{1 -\tau} (1 + \sqrt{1-\tau} ) }
\ .
\end{eqnarray}
The function is well defined and regular (even analytic)
for ${\bf v}^2 = \tau < 1$. The operator
\beq   \label{akl}
\gamma^{-2}{\bf a}^k_{\ l} :=  {\delta}^k_l +
\varphi v^k v_l
\eeq
 acting on rest-frame
variables $x^l$ comes from the boost transformation (for simplicity we skip the
argument ${\bf v}^2$ of the function $\varphi$).

Suppose, therefore, that for a given trajectory $\zeta$ a
{\em local} isometry (\ref{isometry}) has been chosen,
which defines $F_{\zeta} : {\bf M} \mapsto M$. This mapping
is usually not invertible: different points of {\bf M} may
correspond to the same point of space-time $M$ because
different $\Sigma_t$'s may intersect. It enables us,
however, to define the metric tensor on {\bf M} as the
pull-back $F_{\zeta}^*g$ of the Minkowski metric. The
components $g_{\alpha\beta}$ of the above metric are
defined by the derivatives of $F_{\zeta}$, i.~e.~they
depend upon the first and the second derivatives of the
position ${\bf q}(t)$ of our observer.

Because $(x^k)$ are cartesian coordinates on $\Sigma$, the
space-space components of $g$ are trivial: $g_{ij}  =
\delta_{ij}$.  The only non-trivial components of $g$ are,
therefore, the lapse function and the (purely rotational)
shift vector:
\begin{eqnarray}
N & = & \frac 1{\sqrt{-g^{00}}} = \gamma^{-1}
(1+a_ix^i)\ ,
\nonumber \\
N_m &= & g_{0m} =  \gamma^{-1}
\epsilon_{mkl}\omega^k x^l \ ,
\label{shift}
\end{eqnarray}
where $a^i$ is the observer's acceleration vector in the
co-moving frame. The quantity $\omega^m$ is a rotation,
which depends upon the coordination of isometries
(\ref{isometry}) between different $\Sigma_t$'s.  Because
$\omega^m$ depends locally upon the trajectory, it may also
be calculated in terms of the velocity and the acceleration
of the observer, once the identification (\ref{isometry})
has been chosen.  In the case of example (\ref{embedding}),
it is easy to check that
\begin{eqnarray}
a^i = {\bf a}^i_{\ k}
{\dot v}^k  \ ,  \label{ak}
\end{eqnarray}
\begin{eqnarray}       \label{omegam}
{\omega}_m = \om_{ml}\dot{v}^l\ ,
\eeq
where
\beq   \label{omega-ml}
\om_{ml} :=
\gamma
\varphi v^k {\epsilon}_{klm} \ ,
\end{eqnarray}
and ${\dot v}^k$ is the observer's acceleration in the
laboratory frame.

The metric $F_{\zeta}^*g$ is degenerate at singular points
of the identification map (i.e. where the identification is
locally non-invertible because adjacent $\Sigma$'s
intersect, i.~e.~where $N=0$), but this degeneration does
not produce any difficulties in what follows.

\subsection{Lagrango-Hamiltonian}

Once we know the metric (\ref{shift}) on ${\bf M}$, we may
rewrite the invariant Lagrangian density $L$ of the field
theory under consideration, just as in any other
curvilinear system of coordinates.  The Lagrangian obtained
this way depends on the field $\psi$, its first
derivatives, but also on the observer's position, velocity
and acceleration. Variation with respect to $\psi$ produces
field equations in the co-moving system $(x^\alpha)$.  Due
to the relativistic invariance of the theory, variation of
the Lagrangian with respect to the observer's position
${\bf q}$ {\it should not} produce independent equations
but only conservation laws, implied already by the field
equations.

For our purposes we will keep, however, at the same footing
the field degrees of freedom $\psi$ and (at the moment,
physically irrelevant) observer's degrees of freedom $q^k$.
We are going to perform the complete ``Hamiltonization'' of
this theory, i.e. to pass to the Hamiltonian description
both in  field  and  observer's variables.

Let us first perform a partial Legendre transformation, and
pass to the Hamiltonian description of the field degrees of
freedom, keeping the Lagrangian description of the
``mechanical'' degrees of freedom.  For this purpose we
define
\begin{eqnarray}
L_H:=L - \Pi \dot\psi \ , \label{legendre}
\end{eqnarray}
where $\Pi$ is the momentum canonically conjugate to
$\psi$:
\begin{eqnarray}
\Pi := \frac {\partial L}{\partial{\dot\psi}}  \ .
\end{eqnarray}
The function $L_H$ plays the role of a Hamiltonian (with
negative sign) for the fields and a Lagrangian for the
observer's position ${\bf q}$. It is an analog of the {\it
Routhian function} in analytical mechanics (cf.
\cite{Kij-Dar}).  The Lagrango-Hamiltonian $L_H$ generates
the Hamiltonian field evolution with respect to the
accelerated frame, when the ``mechanical degrees of
freedom'' $q^k$ are fixed.  Due to (\ref{shift}), this
evolution is a superposition of the following three
transformations:
\begin{itemize}
\item time-translation in the direction of the local
time-axis of the observer,
\item boost in the direction of the acceleration $a^k$ of
the observer,
\item purely spatial O(3)-rotation $\omega^m$.
\end{itemize}
It is, therefore, obvious that the numerical value of the
generator $L_H$ of such an evolution is equal to
\begin{eqnarray}
L_H = - \gamma^{-1} \left( {\cal H} + a^{k}{\cal
R}_{k} -
\omega^{m}S_{m} \right) \, , \label{elha}
\end{eqnarray}
where ${\cal H} $ is the rest-frame  field energy, ${\cal
R}_k$ is the rest-frame static moment and $S_m$ is the
rest-frame angular momentum, all of them calculated on
$\Sigma$.  The factor $\gamma^{-1}$ in front of
the generator is necessary, because the time $t=x^0$, which
we used to parameterize the observer's trajectory, is not
the proper time along $\zeta$ but the laboratory time.

\subsection{Legendre transformation and relativistic invariance}

 Now, we perform the Legendre transformation also with
respect to the observer's degrees of freedom, and find this
way the complete Hamiltonian of the entire (observer +
field) system. Let us observe that $L_H$ is a 2-nd order
Lagrangian in the observer's variable:
\begin{eqnarray}
L_H = L_H({\bf q},{\bf v},\dot{{\bf v}},\mbox{fields})\ ,
\end{eqnarray}
(for the Hamiltonian description of a theory arising from a
2-nd order Lagrangian see  Appendix~\ref{CanStruc}. A general
discussion of a Hamiltonian formalism arising from  higher order
Lagrangians may be found in \cite{Gitman} - \cite{Nakamura}). Let
$p_k$ and $\pi_k$ denote the momenta canonically conjugated
to $q^k$ and $v^k$ respectively. The phase space of the
entire system, parameterized by $(q^k,v^k,p_k,\pi_k)$ in the
observer's sector and by $(\psi,\Pi)$ in the field sector,
is endowed with the canonical symplectic 2-form:
\begin{eqnarray}
\Omega := dp_k\wedge dq^k + d\pi_k \wedge dv^k +
\Omega^{field}\ ,
\end{eqnarray}
which generates the Poisson bracket for any two observables
$\cal F$ and $\cal G$:
\begin{eqnarray}  \label{Poisson}
\{ {\cal F},{\cal G} \} &:=& \left( \frac{\partial {\cal
F}}{\partial q^k}
\frac{\partial {\cal G}}{\partial p_k} -
 \frac{\partial {\cal F}}{\partial p_k}
\frac{\partial {\cal G}}{\partial q^k} \right) +
\left( \frac{\partial {\cal F}}{\partial v^k}
\frac{\partial {\cal G}}{\partial \pi_k} -
 \frac{\partial {\cal F}}{\partial \pi_k}
\frac{\partial {\cal G}}{\partial v^k} \right)  +
\{ {\cal F},{\cal G} \}^{field}\ .
\end{eqnarray}
Due to the relativistic invariance of the theory the
following observables: $\cal H$, ${\cal R}_k$, $S_m$ and
the rest-frame field momentum ${\cal P}_k$ generate with
respect to (\ref{Poisson}) the Poincar\'e algebra:
\begin{eqnarray}               \label{H-P}
\{{\cal H},{\cal P}_k\}^{field} &=& 0\ ,\nonumber\\
\{{\cal H}, S_k\}^{field} &=& 0\ ,\nonumber\\
\{{\cal H},{\cal R}_k\}^{field} &=& - {\cal P}_k\ ,\nonumber\\
\{{\cal P}_k,{\cal P}_l\}^{field} &=& 0\ ,\nonumber\nonumber\\
\{{\cal P}_k,{\cal R}_l\}^{field} &=& - g_{kl}{\cal H}\ ,\\
\label{Poincare-relations}
\{{\cal P}_k,S_l\}^{field} &=& \epsilon_{klm}{\cal P}^m\ ,\nonumber\\
\{{\cal R}_k,{\cal R}_l\}^{field} &=& -\epsilon_{klm}S^m\ ,\nonumber\\
\{{\cal R}_k,S_l\}^{field} &=& \epsilon_{klm}{\cal R}^m\
   ,\nonumber\\   \label{S-S}
\{S_k,S_l\}^{field} &=& \epsilon_{klm}S^m\ .\nonumber
\end{eqnarray}
To perform the Legendre transformation one has to calculate
$\dot{q}^k$ and $\dot{v}^k$ in terms of $q^k,\ v^k,\ p_k$
and $\pi_k$ from formulae (cf. Appendix~\ref{CanStruc}):
\begin{eqnarray}   \label{Legendre-observer}
p_k = \frac{\partial L_H}{\partial v^k} - \dot{\pi}_k
\ ,\ \ \ \ \ \ \ \ \pi_k =
\frac{\partial L_H}{\partial
\dot{v}^k}\ .
\end{eqnarray}
Since $L_H$ is linear in $\dot{v}$ the Legendre
transformation is singular and it gives rise to the
Hamiltonian theory with constraints (see \cite{Dirac}-
\cite{Sundermeyer}).  The {\it primary constraints} follow from
(\ref{Legendre-observer}):
\begin{eqnarray} \label{primary}
\phi^{(1)}_{\ k} := \pi_k -  \frac{\partial L_H}{\partial
\dot{v}^k} =
 \pi_k +
\gamma^{-1} \left(
{\bf a}^l_{\ k} {\cal R}_{l} -
\om_{mk} S^{m} \right)
\approx 0\ .
\end{eqnarray}
By the weak equality symbol ``$\approx$'' we emphasize that
the quantity $\phi^{(1)}_{\ k}$ is numerically restricted
to be zero but does not identically vanish throughout the
phase space. This means in particular, that it has nonzero
Poisson bracket with the canonical variables.
 This way the complete Hamiltonian reads
\begin{eqnarray}  \label{bf-H}
{\bf H} = p_kv^k + \pi_k\dot{v}^k - L_H = p_kv^k +
\gamma^{-1}\,{\cal H} + \dot{v}^k\phi^{(1)}_{\ k}\ .
\end{eqnarray}
The observer's acceleration $\dot{{\bf v}}$ plays in
(\ref{bf-H}) the role of Lagrangian multiplier.   To
reduce the theory with respect to the constraints
(\ref{primary}) and to find the ``true'' degrees of freedom
one has to find the complete hierarchy of constraints via
so called Dirac-Bergmann procedure (cf. \cite{Dirac} -
\cite{Sundermeyer}).
\begin{PR}
 The
secondary constraints $\phi^{(2)}_{\ k}$ read:
\begin{eqnarray}  \label{secondary}
\phi^{(2)}_{\ k} := \{\phi^{(1)}_{\ k},{\bf H}\} =
- p_{k} + \gamma v_{k}{\cal H} +
\gamma^{-2}{\bf a}^l_{\ k} {\cal P}_{l} \approx 0\ .
\end{eqnarray}
\end{PR}
Let us find the constraint algebra, i.e. ``commutation
relations'' between  constraints $\phi^{(a)}_{\ k}$.
\begin{PR}
\begin{eqnarray}      \label{first-class}
\{ \phi^{(a)}_{\ k} , \phi^{(b)}_{\ l} \} = 0\ ,
\end{eqnarray}
for $a,b=1,2$ and $k,l=1,2,3$.
\end{PR}
  It means that
$\phi^{(1)}_{\ k}$ and $\phi^{(2)}_{\ k}$ are so-called
{\it first-class constraints} (see
\cite{Dirac} - \cite{Sundermeyer}).
One has to continue this algorithm and look for the
constraints which are the ``conservation laws'' for the
$\phi^{(2)}_{\ k}$, i.e.
\begin{eqnarray}
\phi^{(3)}_{\ k} := \{\phi^{(2)}_{\ k},{\bf H}\} \approx 0\ .
\end{eqnarray}
However, the {\it tetriary constraints} $\phi^{(3)}_{\ k}$
are satisfied identically due to the Poincar\`e relations
(\ref{Poincare-relations}).

This way we show that the hierarchy of constraints ends on
the level of secondary constraints and we obtain the
Hamiltonian theory with 6 first-class constraints
$\phi^{(a)}_{\ k}$.
Let $\overline{{\cal P}}$ denotes the constraint subspace
of the phase space $\cal P$, i.e.
\begin{eqnarray}
\overline{{\cal P}} := \{ x \in {\cal P}\ \mid\
\phi^{(a)}_{\ k}(x)=0\ \mbox{for}
\ a=1,2;\ k=1,2,3\}\ ,\nonumber
\end{eqnarray}
and let $e:\overline{{\cal P}}\rightarrow {\cal P}$ be an
embedding. Then, due to (\ref{first-class}), the pull-back
$e^*\Omega$ of the symplectic form $\Omega$ is degenerate
and to each functional  $\phi^{(a)}_{\ k}$ corresponds a
``gauge direction'' $X^{(a)}_{\ k}$, i.e.
\begin{eqnarray}
X^{(a)}_{\ k} \con e^*\Omega = 0\ ,\nonumber
\end{eqnarray}
such that
\begin{eqnarray}
e_*X^{(a)}_{\ k} \con \Omega = \delta\phi^{(a)}_{\ k}\ ,\nonumber
\end{eqnarray}
i.e.  $X^{(a)}_{\ k}$ is a Hamiltonian vector field
corresponding to the functionial $\phi^{(a)}_{\ k}$
($X^{(a)}_{\ k}$ is tangent to $\overline{{\cal P}}$ due to
(\ref{first-class})).

Therefore, our theory is a gauge theory with 6 gauge
parameters: {\bf q} and {\bf v}. Canonically conjugated
momenta {\bf p} and $\mbox{\boldmath $\pi$}$ are subjected
to constraints $\phi^{(a)}_{\ k}$. Thus, reducing the
theory with respect to these constraints (i.e. passing to
gauge equivalence classes) we end up with  the ``true''
degrees of freedom, namely those describing the field.
Fixing the trajectory plays the role of ``gauge fixing''
and the ``evolution equations'' of the observer are
automatically satisfied if the field equations are
satisfied.

This result is an obvious consequence of the relativistic invariance of the
theory. Observer's parameters are not true degrees of freedom and can be easily
``gauged away''.

\section{Renormalized electrodynamical Routhian}
\label{Renorm}
\setcounter{equation}{0}

We would like to apply the formalism presented in the previous Section
for the Hamiltonian
description of electrodynamics of a point particle.
It turns out that the use of the particle's rest-frame simplifies considerably
the formulation of the theory. Hence we will use not an arbitrary observer, but
the one following exactly the particle's trajectory. As a consequence, the
trajectory will no longer be a gauge parameter, but will have an independent,
dynamical meaning, as a new degree of freedom of the theory.

 Let us observe that
starting from the Lagrangian descripton of electrodynamics
given by the standard Maxwell Lagrangian, the Legendre
transformation (\ref{legendre}) does not lead to the
correct local expression for the field energy (we obtain the
``canonical Hamiltonian'' which differs from the field
energy by a complete divergence). However, in the ``field sector'' we may
already start
with the  electrodynamical Routhian (Lagrango-Hamiltonian)
(\ref{elha}). To obtain this generator we take as
${\cal H}$, ${\cal R}_k$ and $S_m$
the conventional energy, static moment and angular momentum
of the electromagnetic field. These quantities are defined
as appropriate integrals of the components of the symmetric
energy-momentum tensor:
\begin{eqnarray}   \label{T}
T^\mu_{\ \nu} = f^{\mu\lambda}f_{\nu\lambda} -
\frac{1}{4}\delta^\mu_{\
\nu}f^{\kappa\lambda}f_{\kappa\lambda}\ .\nonumber
\end{eqnarray}
 In the Section~\ref{Fluxes} we construct the Hamiltonian
structure of Maxwell electrodynamics which is perfectly
suited for the formalism introduced in the previous Section. In
particular, the ``new Hamiltonian'' is positive definite
and equals to the field energy, i.e. is defined as an
integral of $T^\mu_{\ \nu}$.

When one adds a point particle to the electromagnetic
field, then, obviously, the total field energy is not well
defined due to the singularity of the particle's Coulomb field. Therefore, one
has to perform renormalization.
Decomposing the electric induction field on the
rest-frame surface $\Sigma_t$ into the sum
\begin{eqnarray}
{\bf D}= {\bf D}_0 + \overline{\bf D}\nonumber
\end{eqnarray}
of the Coulomb field
${\bf D}_0 = \frac {e{\bf r}}{4\pi r^3}$
and the remaining part $\overline{\bf D}$, we obtain the
following formulae for the renormalized rest-frame
quantities (see \cite{KIJ} and \cite{Kij-Dar} for details):
\begin{eqnarray}
{\cal H} &=&  m +
\frac{1}{2}\int_{\Sigma} (\overline{{\bf D}}^{2} + {\bf
B}^{2})\,d^3x \ , \label{ReH}  \\
{\cal P}_{l} &=&
\int_{\Sigma} (\overline{{\bf D}} \times {\bf B})_{l}\,d^3x
+ \int_{\Sigma} ({\bf D}_0 \times {\bf B})_{l}\,d^3x  \ ,
\label{ReP} \\
 {\cal R}_{k} &=&
\frac{1}{2}\int_{\Sigma} x_{k}(\overline{\bf D}^{2} + {\bf
B}^{2})\,d^3x + \int_{\Sigma} x_{k}\overline{\bf D}{\bf
D}_0\,d^3x\ ,  \label{ReR} \\
S_{m} &=& \int_{\Sigma}
\epsilon_{mkl}x^{k}(\overline{{\bf D}}
\times {\bf B})^{l}\,d^3x \ .\label{ReS}
\end{eqnarray}
Finally, the renormalized Lagrango-Hamiltonian $L_H$ is
defined as follows: replace in (\ref{elha}) the exact
values, i.~e. calculating for the complete field theory,
  of the quantities $\cal H$,
${\cal R}_k$ and $S_m$ by the above renormalized quantities
(\ref{ReH}), (\ref{ReR}) and (\ref{ReS}), containing only
the external Maxwell field.
It was shown in \cite{Kij-Dar} that the variational
principle applied to the renormalized $L_H$ gives the
Euler-Lagrange equations which are equivalent to the
fundamental equation (\ref{fund}).

\section{The new gauge-invariant Hamiltonian structure of Maxwell
electrodynamics}
\label{Fluxes}
\setcounter{equation}{0}

In field theory, contrary to the classical mechanics, there
is no unique way to represent the field evolution as an
infinite-dimensional Hamiltonian system (see \cite{GRG} and
\cite{Kij2}).  Each such representation is based on a
specific choice of controlling the boundary value of the
field, and corresponds to a specific choice of the
Hamiltonian. This non-uniqueness is implied by the
non-uniqueness of the evolution of the portion of the
field, contained in a finite laboratory $V$. Indeed, the
evolution {\it is not} unique because external devices may
influence the field through the open windows of our
laboratory. To choose the Hamiltonian uniquely, we have to
insulate the laboratory or, at least, to specify the
influence of the external world on it.  One may easily
imagine an unsuccessful insulation, which does not prevent
the external field from penetrating the laboratory.  From
our point of view, an insulation is sufficient if it keeps
under control a complete set of field data on the boundary
$\partial V$ in such a way, that the field evolution
becomes mathematically unique.

For relatively simple theories (e.g. scalar field theory)
the Dirichlet problem may be treated as a privileged one
among all possible mixed (initial value + boundary value)
problems which are well posed. It turns out (see
\cite{GRG}) that the Dirichlet problem leads to the
positive definite Hamiltonian. This means that there is a
natural way to insulate the laboratory $V$ adiabatically
from the external world.  But already in electrodynamics
(and even more in General Relativity) any attempt to define
the field Hamiltonian leads immediately to the question:
how do we {\it really} define our Hamiltonian system?

\subsection{Canonical approach}

One usually starts with a canonical dynamical formula for
electrodynamics (see
\cite{Kij-Tulcz}, \cite{Kij2}):
\begin{eqnarray}    \label{a1}
\int_{V}\dot{{\cal F}}^{k0}\delta A_k -
\dot{A}_k\delta{\cal F}^{k0}  & = &
 - \delta H_V^{can} + \int_{\partial V}{\cal F}^{\nu
\perp}\delta A_{\nu}\ ,
\end{eqnarray}
where $H_V^{can}$ is a ``canonical Hamiltonian'' related
via the Legendre transformation to an electrodynamical
Lagrangian density $\cal L$:
\begin{eqnarray}
H_V^{can} = \int_{V} ({\cal F}^{k0}\dot{A}_k - {\cal L})\ .
\end{eqnarray}
We use standard notation: ${\cal F}^{\mu\nu}$ denotes the
electromagnetic induction tenor-density.  The volume $V$
belongs to the hyperplane $\Sigma_t$ and consists of the
exterior of the sphere $S(r_0)$ (by $\perp$ we denote the
component orthogonal to the boundary).  We stress that the
formula (\ref{a1}) is true  also in the case of nonlinear
electrodynamics. For the linear Maxwell theory $\cal L$ is
the standard Maxwell Lagrangian $L_{Maxwell} =
-\frac{1}{4}{\cal F}^{\mu\nu}f_{\mu\nu}$, where $f_{\mu\nu}
:= \partial_{\mu}A_{\nu} - \partial_{\nu}A_{\mu}$ is the
electromagnetic field tensor. In this case the
electromagnetic induction tensor-density is given by ${\cal
F}^{\mu\nu} := \sqrt{-g}g^{\mu \alpha}g^{\nu
\beta}f_{\alpha \beta}$.

To describe the boundary term it is convenient to use
spherical coordinates $(\xi^a )$, $a=1,2,3$; adapted to
$\partial V$. We choose $\xi^3 = r$ as the radial
coordinate and $(\xi^A)$, $A = 1,2$; as angular
coordinates: $\xi^1 =
\Theta$, $\xi^2 = \varphi$. The Euclidean  metric $g_{ab}$ is
diagonal:
\begin{eqnarray}
g_{33} =  1\, , \hspace{1cm} g_{11} = r^{2}\, ,
\hspace{1cm} g_{22} = r^{2}\sin^{2}\Theta \, ,\nonumber
\end{eqnarray}
and the volume element $\lambda = (\det g_{ab})^{1/2}$ is
equal to $r^2\sin\Theta$. With this notation we have:
\begin{eqnarray}    \label{a2}
\int_{V}\dot{{\cal F}}^{k0}\delta A_k -
\dot{A}_k\delta{\cal F}^{k0}  & = &
 - \delta H_V^{can} +
\int_{\partial V}{\cal F}^{03}\delta A_{0} + {\cal
F}^{B3}\delta A_{B}\ .
\end{eqnarray}
The formula (\ref{a2}) is analogous to the Hamiltonian
formula in classical mechanics
\begin{eqnarray}
\dot{p}_kdq^k - \dot{q}^kdp_k = - d H(q,p) \ .\nonumber
\end{eqnarray}
But there is also a boundary term in (\ref{a2}), typical
for field theory.  Killing this term by an appropriate
choice of boundary conditions is necessary for transforming
the field theory into an (infinite dimensional) dynamical
system (see
\cite{Kij-Tulcz}, \cite{GRG}).  Thus, boundary conditions for
$A_0|\partial V$ and $A_B|\partial V$ make the
electrodynamics equivalent to the infinite-dimensional
Hamiltonian system.  From the mathematical point of view
this is the missing part of the definition of the
functional space. The Hamiltonian structure of Maxwell
electrodynamics described above is mathematically well
defined, i.e. a mixed Cauchy problem (Cauchy data given on
$\Sigma_t$ and Dirichlet data given on $\partial V\times
{\bf R}$) has an unique solution (modulo gauge
tranformations which reduce to the identity on $\partial
V\times {\bf R}$).

\subsection{The new approach}

It turns out that there is another Hamiltonian structure
which is also mathematically well defined (see \cite{Kij2},
\cite{GRG}). Moreover, it possesses very nice properties
from the physical point of view:
\begin{enumerate}
\item the new Hamiltonian corresponds to the field energy
obtained from the symmetric energy-momentum tensor, i.e
$\frac{1}{2}(D^2 + B^2)$ in the laboratory frame, which, contrary to the
canonical Hamiltonian, is positive
definite,
\item this structure is perfectly suited for the reduction
of the theory with respect to the Hamiltonian constraint
$D^k,_k=0$ , i.e.  Gauss law (see
\cite{Kij2}).
\end{enumerate}

Let us start from the canonical relation (\ref{a2})
\begin{eqnarray}    \label{a3}
\lefteqn{
\int_{V}\dot{{\cal F}}^{30}\delta A_k + \dot{{\cal F}}^{B0}
\delta A_B -
\dot{A}_k\delta{\cal F}^{30} - \dot{A}_B\delta{\cal F}^{B0}}
\nonumber\\ & = &
- \delta H_V^{can} +
\int_{V}\dot{{\cal F}}^{30}\delta A_3 + \dot{{\cal F}}^{B0}
\delta A_B -
\dot{A}_3\delta{\cal F}^{30} - \dot{A}_B\delta{\cal F}^{B0} \ .
\end{eqnarray}
On each sphere $S(r) = \{r=const\}$ the 2-dimensional
covector field $A_B$ splits into a sum of the
``longitudinal'' and the ``transversal'' part:
\begin{eqnarray}   \label{spliting}
A_{B} = u,_{B} + \epsilon_{B}^{\ \ C}v,_{C}\, ,
\end{eqnarray}
where the coma denotes partial differentiation and
$\epsilon^{AB}$ is a sqew-symmetric tensor, such that
$\lambda\epsilon^{AB}$ is equal to the Levi-Civita
tensor-density (i.~e.~$\lambda\epsilon^{12} =
-\lambda\epsilon^{21} =1$).  The functions $u$ and $v$ are
uniquely given by the field $A_B$ up to additive constants
on each sphere separately. Inserting this decomposition
into (\ref{a3}) and integrating by parts we obtain:
\begin{eqnarray}    \label{a4}
\int_{V}(\dot{{\cal F}}^{30}\delta A_3 - \dot{A}_3
\delta{\cal F}^{30}) - ( \dot{{\cal F}}^{B0},_{B}\delta u
- \dot{u} \delta{\cal F}^{B0},_B) - ( \dot{{\cal
F}}^{B0},_{C}\epsilon_B^{\ C}\delta v  - \dot{v} {\cal
F}^{B0},_{C}\epsilon_B^{\ C} )\nonumber\\
 = - \delta
H_V^{can} + \int_{\partial V}{\cal F}^{03}\delta A_{0} -
{\cal F}^{B3},_{B}\delta u - {\cal
F}^{B3},_{C}\epsilon_B^{\ C}\delta v\  .
\end{eqnarray}
Using identities $\partial_{B}{\cal F}^{B0} +
\partial_{3}{\cal F}^{30} = 0$ and $\partial_{B}{\cal
F}^{B3} + \partial_{0}{\cal F}^{03} = 0$, implied by the
field equations $\partial_{\mu}{\cal F}^{\mu\nu}=0$, and
integrating again by parts we finally obtain:
\begin{eqnarray}   \label{a5}
\int_{V}[ \dot{{\cal F}}^{30}\delta (A_3 -u,_{3}) - (\dot{A}_3 -
\dot{u},_3) \delta{\cal F}^{30}] +
[ \dot{{\cal F}}^{0B||C}\epsilon_{BC}\delta v  -
\dot{v}\delta {\cal F}^{0B||C}\epsilon_{BC} ] \nonumber\\
= - \delta H_V^{can} + \int_{\partial V}{\cal
F}^{03}\delta (A_{0} - \dot{u}) + {\cal
F}^{3B||C}\epsilon_{BC}\delta v\ .
\end{eqnarray}
Here, by ``$||$''we denote the 2-dimensional covariant
derivative on each sphere $S(r)$. The quantities $(A_{0} -
u,_{0})$ and $(A_{3} - u,_{3})$ are ``almost'' gauge
invariant: only their monopole part (mean-value) on each
sphere may be affected if we change the additive constant
in the definition of $u$ (the choice of an additive
constant in the definition of $v$ is irrelevant, because it
is always multiplied by quantities which vanish when
integrated over a sphere). The sum of the volume and
surface integrals in (\ref{a5}) is however gauge invariant.
Now,
\begin{eqnarray}
B^A = ({\bf curl A})^A = \epsilon^{AB}(A_B,_3 - A_3,_B)
\end{eqnarray}
and using (\ref{spliting}) we have
\begin{eqnarray}
\triangle (A_{3} - u_{,3}) = r^{2}B^{A||B}\epsilon_{AB} \, ,
\end{eqnarray}
where $\triangle$ denotes the 2-dimensional
Laplace-Beltrami operator on $S(r)$ multiplied by $r^2$
(the operator $\triangle$ does not depend on $r$ and is
equal to the Laplace-Beltrami operator on the unit sphere
$S(1)$). The operator $\triangle$ is invertible on the
space of monopole--free functions (functions with vanishing
mean value on each $S(r)$). This functional space will play
an important role in further considerations and all the
dynamical field quantities of the theory will belong to
this space. To fix both terms in (\ref{a5}) uniquely we
choose $u$ in such a way that the mean value of $(A_{3} -
u_{,3})$ vanishes on each sphere.  Hence, with the above
choice of the additive constants the quantity $A_{3} -
u_{,3}$ becomes gauge invariant:
\begin{eqnarray} \label{69}
A_{3} - u_{,3} = r^{2}\triangle^{-1}(B^{A||B}\epsilon_{AB})
\, .
\end{eqnarray}

Let us observe that the function $v$ is also gauge
invariant (up to an additive constant, which does not play
any role and may also be chosen in such a way that its mean
value vanishes on each sphere).  Indeed, we have:
\begin{eqnarray}    \label{b3v}
B^{3} = ({\bf curl\, A})^{3} = A^{A||B}\epsilon_{BA} = -
r^{-2}\triangle v\, .
\end{eqnarray}
Due to the Maxwell equation ${\bf div\, B} = 0$, the
function $B^3$ is monopole--free and the Laplasian
$\triangle$ may again be inverted:
\begin{eqnarray} \label{71}
v = - r^{2}\triangle^{-1}B^{3} \, .
\end{eqnarray}
The formula (\ref{a5}) could be also obtained directly from
(\ref{a3}) by imposing the following gauge conditions:
\begin{eqnarray}    \label{gauge1}
A^{B}_{\ ||B} &=& 0 \, ,\\   \label{gauge2}
\int_{S(r)}\lambda A_{3} &=& 0 \, .
\end{eqnarray}
The above condition does not fix the gauge uniquely: we
still may add to $A_\mu$ the gradient of a function of time
$f=f(t)$. This residual gauge changes only the monopole
part of $A_0$, but both the volume and the surface
integrals in (\ref{a2}) remain invariant with respect to
such a transformation.

Assuming the above gauge, we have $u \equiv 0$ and $A_{B} =
\epsilon_{B}^{\ \ C}v,_{C} $. To simplify the notation we
will, therefore, replace our invariants $(A_{0} - u,_{0})$
and $(A_{3} - u,_{3})$ by the values of $A_{0}$ and
$A_{3}$, calculated in this particular gauge.

Let us observe that formula (\ref{a5}) represents dynamical
system with infinitely many degrees of freedom described by
four functions: ${\cal F}^{30}$, $A_{3}$, ${\cal
F}^{B0||C}\epsilon_{BC}$ and $v$ (contrary to (\ref{a2})
described by six functions $A_k$ and $D^k$). Two of them
will play the role of field configurations and the
remaining two will be the conjugate momenta. Let us
consider a boundary term in (\ref{a5}).  Killing this term
by an appropriate choice of boundary conditions is
necessary for transforming the field theory into an
(infinite dimensional) dynamical system (see
\cite{Kij-Tulcz}, \cite{GRG}).
>From this point of view, the quantity $v$ (or, equivalently
$B^3$) is a good candidate for the field configuration,
since controlling it at the boundary will kill the term
$\delta v$ in the boundary integral. On the contrary,
$\delta A_0$ can not be killed by any simple boundary
condition imposed on $A_3$. We conclude, that it is rather
${\cal F}^{03} = \lambda D^3$ than $A_3$, which has to be
chosen as another field configuration.

Hence, we perform the Legendre transformation in formula
(\ref{a5}) on the boundary $\partial V$: ${\cal
F}^{03}\delta A_0 = \delta({\cal F}^{03}A_0) - A_0\delta
{\cal F}^{03}$.  This way, using (\ref{69}) and (\ref{71}),
we obtain from (\ref{a5}) the following result:
\begin{eqnarray}   \label{a6}
\lefteqn{
\int_{V} \lambda r^{2} \left[ \rule{0ex}{3ex} -\dot{D}^3\delta
(\triangle^{-1}(B^{A||B}\epsilon_{AB})) +
\triangle^{-1}(\dot{B}^{A||B}\epsilon_{AB})
\delta D^3  \right.} \nonumber\\ &+& \left. \dot{B}^3\delta
(\triangle^{-1}(D^{A||B}\epsilon_{AB})) -
\triangle^{-1}(\dot{D}^{A||B}\epsilon_{AB})
\delta B^3 \rule{0ex}{3ex} \right]\nonumber\\
&=& \delta \left( H_V^{can} - \int_{\partial V}{\cal
F}^{03}A_0
\right) + \int_{\partial V} - \lambda {A}_{0} \delta D^3 -
r^{2}\triangle^{-1}({\cal F}^{3A||B}\epsilon_{AB})\delta
B^{3}  \  ,
\end{eqnarray}
where we used the fact that the operator $\triangle^{-1}$
is self-adjoint on the functional space of monopole-free
functions on a sphere.

We see that $(D^3 , B^3 )$ play the role of field
configurations, whereas the remaining functions
$(B^{A||B}\epsilon_{AB} ,D^{A||B}\epsilon_{AB})$ describe
the conjugate momenta. Controlling the configurations at
the boundary we kill the surface integral over ${\partial
V}$ and obtain this way an infinite dimensional Hamiltonian
system describing the field evolution. There is, however, a
problem with such a control, because the electric induction
$D^3$ cannot be controlled freely on the boundary. The
reason is that the total electric flux through both
components of $\partial V$ (i.~e.~through $S(r_0 )$ and
through the sphere at infinity) must be the same:
\begin{eqnarray}   \label{constr}
\int_{S(r_{0})}{\cal F}^{03} = \int_{S(r_{\infty})}{\cal F}^{03} = e
\, ,
\end{eqnarray}
where $e$ is the electric charge contained in $S(r_0)$.
Hence, we have to separate the monopole- free
(``radiative'') part of $D^3$ (which can be independently
controlled on both ends of $V$) from the information about
the electric charge.  For this purpose we split the
electric induction $D^3$ into
\begin{eqnarray}   \label{splitD}
D^3 = \frac{e}{4\pi r^2} + \overline{D}^3\, ,
\end{eqnarray}
where $\overline{D}^3$ is again a monopole-free function.
It follows from (\ref{constr}) that the monopole part of
$D^3$ (equal to $e/4\pi r^2$) is nondynamical and drops out
from the volume integral of (\ref{a3}) because it is
multiplied by a monopole-free function
$B^{A||B}\epsilon_{AB}$. The remaining part
$\overline{D}^3$ (which does not carry any information
about the charge $e$),  together with $B^3$, can be taken
as the true, unconstrained degrees of freedom of the
electromagnetic field.

In the same way we split the scalar potential $A_0$:
\begin{eqnarray}   \label{splitA}
A_0 = \phi + \overline{A}_0\, ,
\end{eqnarray}
where $\phi(r)$ is the mean value of $A_0$ on the sphere
$S(r)$ (monopole part) and $\overline{A}_0$ is a
monopole--free function (``radiative'' part of $A_0$).
 Now, the boundary term $A_0\delta D^3$ in
(\ref{a3}) reads
\begin{eqnarray}   \label{A0D3}
\int_{\partial V} \lambda A_0\delta D^3 = \frac{1}{4\pi}
\int_{\partial V}
\lambda r^{-2} \phi \delta e + \int_{\partial V} \lambda
\overline{A}_0
\delta \overline{D}^3\ .
\end{eqnarray}
Finally, we perform the Legendre transformation between
${\phi}$ and the monopole part of $D^3$ at infinity. Hence,
we control the total charge contained in $S(r_0)$ and the
monopole function ${\phi}$ at infinity. Since the latter
does not contain any physical information and is used only
to fix the residual gauge, we may use the simplest possible
choice: ${\phi} (\infty ) \equiv 0$. This way we have proved the following
\begin{TH}
The quantities
\begin{eqnarray*}
\Psi^{1} & = & rB^{3}\, ,\\
\Psi^{2} & = & r\overline{D}^3\, ,\\
\chi_{1} & = & - r\triangle^{-1}(D^{A||B}\epsilon_{AB})\, ,\\
\chi_{2} & = & r\triangle^{-1}(B^{A||B}\epsilon_{AB})\,
\end{eqnarray*}
together with the value $e$ of the electric charge
contained in $S(r_0 )$ contain the entire (gauge
invariant) information about the electromagnetic field.
\end{TH}
Quantities $\Psi^A$ play the role of  field configurations
and $\chi_A$ are conjugated momenta.
Finally, formula (\ref{a6}) reads:
\begin{eqnarray}   \label{Hamiltonian-formula}
\int_{V}\lambda (\dot{\chi}_{A}\delta \Psi^{A}-
\dot{\Psi}^A\delta\chi_A) &=& - \delta {H}_V +
\int_{\partial V}\lambda\chi_{A}^{r}\delta \Psi^{A} +
e\delta{\phi}(\infty) + {\phi}(r_0)\delta e
\end{eqnarray}
where the new Hamiltonian ${H}_V$ equals
\begin{eqnarray}   \label{ham}
{H}_V = H_V^{can} - \int_{\partial V}{\cal F}^{03}A_0\ ,
\end{eqnarray}
and the ``boundary momenta'' are given by:
\begin{eqnarray*}  \label{bm1}
\chi_{1}^{r} & = & - r\triangle^{-1}(\frac{1}{\lambda}{\cal
F}^{3A||B}\epsilon_{AB})\, ,\\   \label{bm2}
\chi_{2}^{r} & = & - r^{-1}\overline{A}_0\, .
\end{eqnarray*}
They describe the response of the system to the control of the boundary
values of  configurations $\Psi^A$.
The above Hamiltonian corresponds to the symmetric
energy-momentum tensor of the Maxwell field (cf.
\cite{GRG}, \cite{Kij2}), i.e. $H_V$ equals numerically to
the amount of the electromagnetic energy contained in the
volume $V$.  Obviously, the limit $\lim_{r_0\rightarrow
0}H_V$ is not well defined due to the Coulomb field
singularity. However,  renormalizing it one gets
exactly (up to a sign) the renormalized
Lagrango-Hamiltonian $L_H$.

The space of the electromagnetic field contained in $V$
$${\cal P}^{field}_{r_0} = \{\Psi^A,\chi_A:V\rightarrow
{\bf R}\ |\
\mbox{with boundary conditions}\ \Psi^A|\partial V \}$$
is endowed with the canonical symplectic 2-form
\begin{eqnarray}  \label{Omega-Maxwell}
\Omega^{field}_{r_0} := \int_{V}\lambda \delta\chi_A
\wedge \delta\Psi^A \ .
\end{eqnarray}
It may be easily obtained (see \cite{GRG}) by the reduction
of the standard presymplectic form $ \int_{V} \delta{\cal
F}^{k0} \wedge \delta A_k $ with respect to the Hamiltonian
constraint $\partial_k{\cal F}^{k0}=0$.  The form
$\Omega^{field}_{r_0}$ defines in the space of physical
observables, i.e. functionals over ${\cal
P}^{field}_{r_0}$, the canonical Poisson bracket structure:
\begin{eqnarray}    \label{Poisson-Maxwell}
\{{\cal F},{\cal G}\}^{field}_{r_0} := \int_{V}\lambda \left(
\frac{\delta {\cal F}}{\delta{\Psi^A}({\bf x})}
\frac{\delta {\cal G}}{\delta{\chi_A}({\bf x})} -
\frac{\delta {\cal G}}{\delta{\Psi^A}({\bf x})}
\frac{\delta {\cal F}}{\delta{\chi_A}({\bf x})} \right)\ .
\end{eqnarray}

\section{Reduction }
\label{Reduction}
\setcounter{equation}{0}

In this Section we finally find the space of Hamiltonian
variables and the total Hamiltonian for
electrodynamics of a point particle. We start with the renormalized
Lagrango-Hamiltonian. It generates the Hamiltonian dynamics in the ``field
sector'' and Lagrangian dynamics in the ``particle's sector''.
 Our aim is to perform
complete ``Hamiltonization'' of the theory, i.e. to perform
the Legendre transformation in the particle's variables.

The original phase space is a direct sum of a particle's
space ${\cal P}^{particle}$ and a phase space of the
Maxwell field ${\cal P}^{field}$:
\begin{eqnarray}
   {\cal P} = {\cal P}^{particle} \oplus {\cal P}^{field}\ ,\nonumber
\end{eqnarray}
where
\begin{eqnarray}
{\cal P}^{field} := \lim_{r_0\rightarrow 0} {\cal
P}^{field}_{r_0}\ .\nonumber
\end{eqnarray}
The particle's phase space ${\cal P}^{particle}$ is
parameterized by $q^k,\ v^k$ and conjugated momenta $p_k$,
$\pi_k$ and ${\cal P}^{field}$ by $\Psi^A$ and conjugated
momenta $\chi_A$ with the boundary condition
$\Psi^A|\partial V$ (now $V$ is an exterior of the sphere
$S(r_0)$ with $r_0 \rightarrow 0$).
The total  phase space $\cal P$ is endowed with the
canonical symplectic form $\Omega$ given by
\begin{eqnarray}  \label{Omega1}
\Omega := dp_k\wedge dq^k + d\pi_k \wedge dv^k + \Omega^{field}\ ,
\end{eqnarray}
 which generates the Poisson bracket in the space of
functionals over $\cal P$:
\begin{eqnarray}  \label{Poisson1}
\{ {\cal F},{\cal G} \} &:=& \left( \frac{\partial {\cal
F}}{\partial q^k}
\frac{\partial {\cal G}}{\partial p_k} -
 \frac{\partial {\cal F}}{\partial p_k}
\frac{\partial {\cal G}}{\partial q^k} \right) +
\left( \frac{\partial {\cal F}}{\partial v^k}
\frac{\partial {\cal G}}{\partial \pi_k} -
 \frac{\partial {\cal F}}{\partial \pi_k}
\frac{\partial {\cal G}}{\partial v^k} \right)  + \{{\cal
F},{\cal G}\}^{field}\ ,
\end{eqnarray}
where
\begin{eqnarray}
\Omega^{field} &:=& \lim_{r_0\rightarrow 0}
\Omega^{field}_{r_0}\ ,\\
\{{\cal F},{\cal G}\}^{field} &:=& \lim_{r_0\rightarrow 0}
\{{\cal F},{\cal G}\}^{field}_{r_0}\ ,
\end{eqnarray}
and $\{\ ,\ \}^{field}_{r_0}$ is given by (\ref{Poisson-Maxwell}).
In the case of complete field theory the rest frame functionals: $\cal H$,
${\cal P}_k$, ${\cal R}_k$ and ${\cal S}_m$ form the Poincar\'e algebra with
respect to $\{\ ,\ \}^{field}$. It is no longer true for the renormalized
functionals. This does not mean that renormalized electrodynamics is
incompatible with the special theory of relativity. Notice, that already in the
context of inhomogeneous Maxwell theory with given point-like sources, field
functionals do not form the Poincar\'e algebra but the theory is obviously
relativistically invariant. We have shown in Section~\ref{Comoving}
 that the Poincar\'e
algebra structure is equivalent to the fact that the Legendre transformation
(\ref{legendre}) leads to the Hamiltonian theory with first class constraints. A
theory which is subjected to the first class constraints is a gauge-type
theory. Now, we expect that the particle's trajectory is no longer a gauge
parameter but plays a dynamical role. Therefore, the ``breaking'' of the
Poincar\'e algebra structure is the necessary condition for the nontrivial
particle's dynamics.

To effectively reduced $\cal P$ we shall proceed as follows: we
consider particle's trajectory $\zeta$ as a
limiting case of a tiny world-tube of radius $r_0$.
Therefore,  calculating Poisson brackets according to (\ref{Poisson1})
we shall keep
$r_0 > 0$ and then finally go to the limit $r_0 \rightarrow
0$.

One can easily shows that only three among nine Poincar\'e relations
 (\ref{Poincare-relations})   are
changed, namely:
\beq
\{{\cal P}_k,{\cal H}\}^{field}_{r_0} &=& - e\left(\frac{\alpha_k}{6\pi r_0}
 + \beta_k\right)\ ,\nonumber\\
\{{\cal P}_k,{\cal P}_l\}^{field}_{r_0} &=&
\epsilon_{klm}B^m(r_0)\ ,\\
\{{\cal P}_k,{\cal R}_l\}^{field}_{r_0} &=&
-g_{kl} \left( {\cal H} - m + \frac{e^2}{6\pi r_0} \right) \ ,\nonumber
\eeq
where $B^m(r_0)$ denotes the mean value of the $m$-th component of {\bf B} on
the sphere $S^2(r_0)$. Functions $\alpha_k$ and $\beta_k$ are components of the
dipole parts of the functions $\alpha$ and $\beta$, given by decompositions:
\beq
\mbox{DP}(\alpha) = \alpha_k \frac{x^k}{r}\ ,\ \ \ \ \ \ \ \ \
\mbox{DP}(\beta) = \beta_k \frac{x^k}{r}\ .  \nonumber
\eeq
Functions $\alpha$ and $\beta$ are $r^{-1}$ and $r^0$ terms respectively in the
decomposition of the radial component $D^r$ in the vicinity of the particle
(see (\ref{expansion})).

Let us perform the Legendre transformation in the
particle's sector in the same way as in the
Section~\ref{Comoving}. The renormalized
Lagrango-Hamiltonian gives rise to the Hamiltonian theory
with constraints.  The primary constraints $\phi^{(1)}_{\
k}$ are given obviously by (\ref{primary}). It is easy to show that
the secondary constraints $\phi^{(2)}_{\ k}$
\begin{eqnarray}  \label{secondary1}
\phi^{(2)}_{\ k} := \{\phi^{(1)}_{\ k},{\bf H}\}_{r_0} \ ,
\end{eqnarray}
where the complete Hamiltonian {\bf H} is given by
(\ref{bf-H}), have the same form as in  (\ref{secondary}).
The next step in the Dirac-Bergmann procedure is to calculate the Poisson
brackets between constraints.
\begin{PR}
\begin{eqnarray}   \label{1-1}
\{\phi^{(1)}_{\ k},\phi^{(1)}_{\ l}\}_{r_0} &=& 0\ ,\nonumber\\
\label{1-2}
\{\phi^{(1)}_{\ k},\phi^{(2)}_{\ l}\}_{r_0} &=& \gamma
\left( m - \frac{e^2}{6\pi r_0}\right)\left( g_{kl}
+ \gamma^2 v_kv_l\right)\ ,\\
\label{2-2}
\{\phi^{(2)}_{\ k},\phi^{(2)}_{\ l}\}_{r_0} &=&
2\gamma ev_{[k}\delta^i_{l]}\left(
\frac{\alpha_i}{6\pi r_0} + \beta_i\right) +
e\epsilon_{klm}B^m(r_0)\ ,\nonumber
\end{eqnarray}
where $2A_{[k}B_{l]} := A_kB_l - A_lB_k$.
\end{PR}
We conclude  that the constraints $\phi^{(1)}_{\ k}$ and
$\phi^{(2)}_{\ k}$ are the same as in the case of fundamental
theory, however,  the constraint algebra is completely
different.

\begin{DF}
 A functional $\cal F$ over $\cal P$ is said to be
{\it first-class} if its Poisson bracket with every
constraint vanishes weakly.
\end{DF}
 Observe that due to
(\ref{1-2}) there is no functional among
$\phi^{(a)}_{\ k}$ which is first-class.  The absence of
the first-class constraints means that there are no gauge
parameters in the theory.
The constraints which are not first-class are called {\it
second-class}.  We
conclude that the renormalized Lagrango-Hamiltonian gives
rise to the Hamiltonian theory with second-class
constraints $\phi^{(a)}_{\ k}$.

Let $\overline{{\cal P}}$ denotes the constrained
submanifold of ${\cal P}$, i.e.
\begin{eqnarray}
\overline{{\cal P}} := \{ x\in{\cal P}\ |\ \phi^{(a)}_{\ k}(x)
= 0\}\ ,\nonumber
\end{eqnarray}
and let $e:\overline{{\cal P}}\rightarrow{\cal P}$ be an
embedding. Due to the fact that $\phi^{(a)}_{\ k}$ are
second-class constraints the pull-back $\overline{\Omega} =
e^*\Omega$ of the symplectic form $\Omega$ is a
non-degenerate 2-form on $\overline{{\cal P}}$ (this is
equivalent to the fact that there is no gauge freedom at
all). One may think that $(\overline{{\cal P}},\
\overline{\Omega})$ is an adequate phase space for electrodynamics
of a point particle. But it is not the whole story.

\begin{PR}
In the finite-dimensional case a symplectic form defines an
isomorphism between vectors and covectors (1-forms), i.e.
if $(P,\omega)$ is a symplectic manifold, then $\omega$
induces a continuous linear map for any point $p\in P$:
$\omega^{\flat}_p : T_pP \rightarrow T^*_pP$ defined by
\begin{eqnarray}
\omega^{\flat}_p(X)\cdot Y := \omega_p(X,Y)\ ,\nonumber
\end{eqnarray}
for any $X,Y \in T_pP$. If $\mbox{dim}\,P < \infty$, then
$\omega^{\flat}_p$ is an isomorphism.
\end{PR}
This theorem, however, is no longer true in the infinite-dimensional case.

\begin{PR}
If $\mbox{dim}\,P =
\infty$, then in general $\omega^{\flat}_p$ is only
injective, i.e.  if $\omega^{\flat}_p(X,Y)=0$ for all $Y\in
T_pP$, then $X=0$.
\end{PR}
See \cite{Marsden} for the proof.

\begin{DF}
If $\omega^{\flat}_p$ is only
injective, then
$\omega$ is called a
weak symplectic form.
 If, moreover, $\omega^{\flat}_p$ is onto,
then $\omega$ is called a  strong symplectic.
\end{DF}
Let $(P,\omega)$ be a weak symplectic manifold and let $X$
be a vector field on $P$ defined on a dense subset $D$ of
$P$.
\begin{DF}
We call $X$ a Hamiltonian vector field if there exists
a functional ${\cal F}:D\rightarrow {\bf R}$ such that
\begin{eqnarray}
X \con \omega = - \delta {\cal F}\nonumber
\end{eqnarray}
is satisfied on $D$.
\end{DF}
Let us observe, that for a weak symplectic $\omega$,
 there need not exist a vector field $X_{{\cal
F}}$ corresponding to every given functional $\cal F$ on
$D$. Therefore, from the physical point of view, weak structure
is ``too weak''.

Let us come back to a manifold $\overline{{\cal P}}$ with a
nondegerate 2-form $\overline{\Omega}$. Is $(\overline{{\cal
P}},\overline{\Omega})$ weak or strong?
It turns out that
$\overline{\Omega}$ is only a weak symplectic form on
$\overline{{\cal P}}$ and the reason is very physical. The Hamiltonian vector
field corresponding to the Hamiltonian {\bf H} is not well defined on
$\overline{{\cal P}}$.
To simplify our considerations let us parametrize
$\overline{{\cal P}}$ by a suitable coordinate system. The
simplest parametrization is the following: field sector is
still parameterized by $\Psi^A$ and $\chi_A$ and  particle's
sector by {\bf q} and {\bf v}. Momenta $p_k$ and $\pi_k$
are completely determined by constraints (\ref{primary})
and (\ref{secondary}):
\begin{eqnarray}  \label{pik}
\pi_k({\bf q},{\bf v},\mbox{fields})
 &=& -\gamma^{-1} ({\bf a}^l_{\ k} {\cal R}_l
- \om_{mk}S^m )\ ,\\  \label{pk}
p_k({\bf q},{\bf v},\mbox{fields})
 &=& \gamma v_k{\cal H} + \gamma^{-2}{\bf a}^l_{\ k}{\cal P}_l\ .
\end{eqnarray}
In fact $\pi_k$ and $p_k$ do not depend on {\bf q}.
Notice, that momentum $p_k$ canonically conjugated to
particle's position $q^k$ equals to the total momentum of
the composed system (particle + field) in the laboratory
frame.
Now, using (\ref{bf-H}) and (\ref{pk}) the complete
Hamiltonian {\bf H} on $\overline{{\cal P}}$ is given by
\begin{eqnarray}         \label{H-qv}
{\bf H}({\bf q},{\bf v};\mbox{fields}) =
\gamma \,\left({\cal H} + v^k{\cal
P}_k\right)\ ,
\end{eqnarray}
and equals to the total energy of the composed system in
the laboratory frame.  Obviously, {\bf H} is well defined
at any point of $\overline{{\cal P}}$.  However, the
Hamiltonian vector field $X_{{\bf H}}$ is defined only on a
subset ${\cal P}^*$ of $\overline{{\cal P}}$.
\begin{TH}   \label{TH1}
The Hamiltonian vector field $X_{{\bf H}}$ is well defined
if and only if the ``fundamental equation'' (\ref{fund})
is satisfied, i.e. ${\cal P}^*$ is defined by the following
condition:
\end{TH}
\begin{eqnarray}    \label{FUND}
\mbox{DP} ( 4\pi m\Psi^2 + e^2\Psi^2,_3)(0) = 0\ .\nonumber
\end{eqnarray}
The proof of Theorem~\ref{TH1} is given in the
Appendix~\ref{PROOF}.  The main result of this Section consists in the
following
\begin{TH}   \label{strong}
$({\cal P}^*,\Omega^*)$ is the strong symplectic manifold.
\end{TH}
The proof of this Theorem is given in the next Section
where we construct the reduced Poisson bracket on ${\cal
P}^*$. It turns out that the reduced bracket of any two
well defined functionals over ${\cal P}^*$ is well defined
throughout the reduced phase space.

Therefore, we finally take the space $({\cal
P}^*,\Omega^*)$, where $\Omega^*$ is a reduction of
$\overline{\Omega}$ to ${\cal P}^*$, as a phase space for
the composed system. Due to Theorem~\ref{strong} the
Hamiltonian structure $({\cal P}^*,\Omega^*,{\bf H})$ for
electrodynamics of a point particle is well defined: each
state in ${\cal P}^*$ uniquely determines the entire
history of the system.
Observe, that due to (\ref{pik}) and (\ref{pk}) the reduced
symplectic form possesses highly nontrivial form:
\begin{eqnarray}   \label{Omega*}
\lefteqn{\Omega^*\, =\, \frac{\partial p_k}{\partial
v^l}dv^l\wedge dq^k +
\int_{\Sigma}\lambda\frac{\delta p_k}{\delta \Psi^A}
\delta\Psi^A \wedge dq^k +
\int_{\Sigma}\lambda\frac{\delta p_k}{\delta \chi_A}
\delta\chi_A \wedge dq^k} \nonumber\\
&+& \left( \frac{\partial \pi_k}{\partial v^l} -
\frac{\partial
\pi_l}{\partial v^k}\right) dv^l \wedge dv^k +
\int_{\Sigma}\lambda\frac{\delta \pi_k}{\delta \Psi^A}
\delta\Psi^A \wedge dv^k +
\int_{\Sigma}\lambda\frac{\delta \pi_k}{\delta \chi_A}
\delta\chi_A \wedge dv^k \nonumber\\
&+& \int_{\Sigma}\lambda \delta\chi_A \wedge
\delta\Psi^A\ .
\end{eqnarray}

It was shown in \cite{KIJ} that the ``fundamental
equation'' is equivalent to the total momentum
conservation. In the Lagrangian formulation of the theory
\cite{Kij-Dar} this equation is nothing else that the
Euler-Lagrange equation of the variational problem. Now, in
the Hamiltonian formulation, the ``fundamental equation''
is already present in the definition of the phase space of
the system and the name ``fundamental'' is fully justified.

\section{The Poisson bracket}
\label{Poisson-bracket-structure}
\setcounter{equation}{0}

In this Section we find the Poisson bracket structure for
electrodynamics of a point particle, i.e. we find the
reduced Poisson bracket $\{\ ,\ \}^*$ for the functionals
over ${\cal P}^*$. It is defined in a obvious way via the
symplectic form $\Omega^*$ on ${\cal P}^*$:
\begin{DF}
For any two functionals $\cal F$ and $\cal G$ over ${\cal
P}^*$
\begin{eqnarray}
\{{\cal F},{\cal G}\}^* := \Omega^*(X_{{\cal F}},X_{{\cal G}})\ .
\nonumber
\end{eqnarray}
\end{DF}
 To find the explicite form of $\{\ ,\ \}^*$ let us
apply the Dirac method \cite{Dirac} (that is way some
authors, e.g. \cite{Hanson} - \cite{Sundermeyer},
 call it the Dirac bracket). As
in the previous Section we find it firstly for $r_0>0$ and
then go to the limit $r_0
\rightarrow 0$.

Using constraint algebra (\ref{1-1})-(\ref{2-2}) let us
define a following matrix (so-called Dirac matrix):
\[ C_{ij}(r_0) := \left( \begin{array}{cc}
\{\phi^{(1)},\phi^{(1)}\}_{r_{0}} &
\{\phi^{(1)},\phi^{(2)}\}_{r_{0}} \\
\{\phi^{(2)},\phi^{(1)}\}_{r_{0}} & \{\phi^{(2)},\phi^{(2)}\}_{r_{0}}
\end{array} \right)_{ij}
 = \left( \begin{array}{cc} 0 & X(r_0) \\ - X(r_0) & Y(r_0)
\end{array} \right)_{ij}\, . \]
Notice, that on ${\cal P}^*$ , i.e. on a dense subset of
$\overline{{\cal P}}$ where the ``fundamental equation'' is
satisfied, the constraint algebra reduces to
\begin{eqnarray}
X_{ij}(r_0) &=& \gamma \left( m -
\frac{e^2}{6\pi r_0}\right)\left( g_{ij} +
\gamma^2 v_iv_j\right)\ ,\\
Y_{ij}(r_0) &=&
2\frac{e}{m}\gamma v_{[i}\beta_{j]} \left(m -
\frac{e^2}{6\pi r_0} \right) + e\epsilon_{ijm}B^m(r_0)\ .
\end{eqnarray}
For a Hamiltonian theory with second-class constraints the
Dirac matrix is non-singular  and the
inverse to $C_{ij}(r_0)$ reads:
\[ C^{-1}(r_0) := \left( \begin{array}{cc}
B(r_0) & - A(r_0) \\ A(r_0) & 0
\end{array} \right)  \ ,   \]
where
\begin{eqnarray}
A^{ij}(r_0) & := & (X^{-1}(r_0))^{ij} = \frac{\gamma^{-1}
(g^{ij} -
v^{i}v^{j})}{m-\frac{1}{6\pi}\frac{e^{2}}{r_{0}}}\, ,\\
B^{ij}(r_0) & := & (X^{-1}(r_0)Y(r_0)X(r_0)^{-1})^{ij}
=\nonumber\\ &=& \frac{e}{m}\frac{\gamma^{-3}
v^{[i}\,\beta^{j]}}{m-\frac{1}{6\pi}\frac{e^{2}}
{r_{0}}} + e\gamma^2 B^{m}(r_0)\frac{\epsilon_{m}^{\;\;ij} -
2v^{k}v^{[i}\epsilon^{j]}_{\;\;\;km}}{(m -
\frac{1}{6\pi}\frac{e^{2}}{r_{0}})^{2}}\, .
\end{eqnarray}
Finally, the ``Dirac bracket'' is defined as follows
\begin{eqnarray}           \label{Dirac-bracket}
\{{\cal F},{\cal G}\}^* := \lim_{r_0\rightarrow 0} \{{\cal F},{\cal
G}\}^*_{r_0}\ ,
\end{eqnarray}
where
\begin{eqnarray}
\{{\cal  F},{\cal G}\}^{*}_{r_{0}} & := & \{{\cal F},{\cal
G}\}_{r_{0}}+A^{ij}(r_0) \left(\rule{0ex}{3ex}\{{\cal
F},\phi^{(1)}_{\ i}\}_{r_{0}}\{\phi^{(2)}_{\ j}, {\cal
G}\}_{r_{0}} - \{{\cal F},\phi^{(2)}_{\
i}\}_{r_{0}}\{\phi^{(1)}_{\ j}, {\cal G}\}_{r_{0}}\right)
\nonumber\\ & - &  B^{ij}(r_0)\{{\cal F},\phi^{(1)}_{\
i}\}_{r_{0}}\{\phi^{(1)}_{\ j}, {\cal G}\}_{r_{0}}\, .
\end{eqnarray}
We stress that the reduced Poisson bracket $\{\ ,\ \}^*$ is
well defined for any two well defined functionals $\cal F$
and $\cal G$ on the reduced phase space ${\cal P}^*$.
Moreover, particle's and field degrees of freedom are
kept at the same footing. To our knowledge it is the
first consistent Poisson bracket structure for the theory
of interacting particles and fields.

The complete set of ``commutation relations'' defined by
$\{\ ,\ \}^*$ is given in the
Appendix~\ref{Commutation-relations}.  Using these
relations one easily prove the following
\begin{TH}
Hamilton equations:
\begin{eqnarray*} \label{nawH1}
\frac{d}{dt}\,\Psi^{A}({\bf x}) & = & \{\Psi^{A}({\bf
x}),{\bf H}\}^{*}\ ,\\
\label{nawH2}
\frac{d}{dt}\,\chi_{A}({\bf x}) & = & \{\chi_{A}({\bf
x}),{\bf H}\}^{*}\ ,
\end{eqnarray*}
reconstruct Maxwell equations in the co-moving frame (cf. \cite{Kij-Dar}).
 Moreover,
\beq
\dot{q}^{k} = \{q^{k},{\bf H}\}^{*} = v^{k}\ ,\nonumber
\eeq             and
\beq
\dot{v}^{k} = \{v^{k},{\bf H}\}^{*} = \frac em
({\bf a}^{-1})^k_{\ l}\beta_l\ ,\nonumber
\eeq
where the inverse of the operator {\bf a} defined in  (\ref{akl}) reads
\beq    \label{a-1}
({\bf a}^{-1})^k_{\ l} := \gamma^{-2}
 \left( \delta^k_{\ l} -
\gamma^{-1} \varphi v^{k}v_{l}\right)\ , \nonumber
\eeq
  gives the particle's
``equation of motion'':
\begin{eqnarray}
ma_k = e\beta_k\ .\nonumber
\end{eqnarray}
\end{TH}

\section{The Hamiltonian}
\label{Hamiltonian}
\setcounter{equation}{0}

The formula (\ref{H-qv}) gives the quasi-local Hamiltonian  for the composed
``particle + field'' system. It is expressed in therms of {\bf q}, {\bf v} and
field variables. However, these
variables  are highly
noncanonical with respect to $\{\ ,\ \}^*$.  Let us observe
that momentum {\bf p} and the particle's position {\bf q}
are still conjugated to each other with respect to the reduced
bracket (\ref{Dirac-bracket}). Indeed, one easily shows
that
\begin{eqnarray}
\{q^i,p_j\}^* = \delta^i_{\ j}\ .\nonumber
\end{eqnarray}
Therefore, we can use {\bf p} instead of {\bf v} to
parametrize ${\cal P}^*$.  However, to express the complete
Hamiltonian {\bf H} in terms of {\bf p} one has to express
{\bf v} in terms of {\bf q}, {\bf p} and fields. Using
(\ref{pk}) one finds rather complicated formula for the
particle's velocity (see Appendix~\ref{velocity-momentum}):
\begin{eqnarray}  \label{vk}
v_{k} = \frac{p^{l}(p_{l} - {\cal P}_{l})\sqrt{{\cal H}^{2}
+ p^{k} p_{k} - {\cal P}^{k}{\cal P}_{k}} - {\cal H}\,{\cal
P}^{l}(p_{l} - {\cal P}_{l})}{[p^{l}(p_{l} - {\cal
P}_{l})]^{2} + {\cal H}^{2}(p^{l} - {\cal P}^{l})(p_{l} -
{\cal P}_{l})}(p_{k} - {\cal P}_{k})\  .
\end{eqnarray}
Now, inserting (\ref{vk}) into ${\bf H} = p_kv^k +
\gamma^{-1}{\cal H}\ $ we obtain
\begin{eqnarray}   \label{H-qp}
{\bf H}({\bf q},{\bf p};\mbox{fields}) = \sqrt{{\cal H}^{2}
+ {\bf p}^2 -
\mbox{\boldmath   $\cal P$}^2 }\ .
\end{eqnarray}
Let us observe that for a free particle, i.e. $e = 0$,
${\cal H} = m$ and ${\cal P}_k = 0$, formula (\ref{vk})
gives relativistic relation between particle's velocity and
momentum:
\begin{eqnarray}
v_{k} = \frac{p_{k}}{\sqrt{m^{2} + {\bf p}^{2}}} \  ,\nonumber
\end{eqnarray}
and formula (\ref{H-qp}) gives relativistic particle's
energy:
\begin{eqnarray}
E({\bf p}) = \sqrt{m^{2} + {\bf p}^{2}}\,  .\nonumber
\end{eqnarray}

\section{Poincar\'e algebra}
\label{Poincare}
\setcounter{equation}{0}

Let us define the laboratory-frame Poincar\'e generators.
They are given by the Lorentz transformation of the
rest-frame generators ${\cal P}^\mu := ({\cal H}, {\cal
P}^k),\ {\cal R}_k$ and $S^m$.  Obviously,
laboratory-frame four-momentum $p^\mu$ is given by
(\ref{pk}) and (\ref{H-qv}):
\begin{eqnarray*}
p^0 &:=& \gamma \left({\cal H} +
v^k{\cal P}_k\right)\ ,\\ p_k &:=& \gamma v_k\,{\cal H} + \gamma^{-2}{\bf
a}^l_{\ k}
{\cal P}_l\ .
\end{eqnarray*}
The static moment $r_k$ and the angular-momentum $s^m$ with
respect to the particle's position ${\bf q}=(q^k)$ are
given as follows:
\begin{eqnarray*}
r_k &:=& \gamma^3 ({\bf a}^{-1})^l_{\ k}{\cal R}_l
+ \gamma \epsilon_{klm}v^lS^m +
q_kp^0\ ,\\ s^m &:=& \gamma^3 ({\bf a}^{-1})^m_{\ k}S^k
- \gamma \epsilon^{mkl}v_k{\cal R}_l
+
\epsilon^{mkl}q_kp_l\ .
\end{eqnarray*}
With these definitions one can prove the following
\begin{TH}
Laboratory-frame generators $p^0,\ p_k,\ r_k$  and $s^m$
form with respect to the reduced Poisson bracket $\{\ ,\
\}^*$ the Poincar\'e algebra.
\end{TH}
This already proves that our formulation is perfectly consistent with the
Lorentz invariance of the theory.

\section{Particle in an external potential}    \label{External}
\setcounter{equation}{0}

Suppose now that the particle moves in an external
(generalized) potential $U = U({\bf q},\dot{\bf q},t)$.
Then the Lagrango-Hamiltonian is given by:
\begin{eqnarray}     \label{LagU}
L_H = - \gamma^{-1}\left({\cal H} +
a^k {\cal R}_k -
\omega_mS^m \right) - U\ .
\end{eqnarray}
To pass into Hamiltonian description in the particle's
variables one has to apply the same procedure as in the
Section~\ref{Reduction}. The primary constraints
$\phi^{(1)}_{\ k}$ are obviously given by (\ref{primary})
but the secondary ones read
\begin{eqnarray}   \label{secondaryext}
\phi^{(2)}_{\ k} =- p_k + \gamma v_k {\cal H} +
\gamma^2 {\bf a}^l_{\ k} {\cal P}_l -
\frac{\partial U}{\partial \dot{q}^k} \approx 0\ .
\end{eqnarray}
Obviously, $\phi^{(a)}_{\ k}\ \mbox{for}\ a=1,2;\ k=1,2,3$
are second-class constraints and the constraint algebra has
the same form as (\ref{1-2}) with $\beta_k$
replaced by $\beta_k^{total}$ given by
\begin{eqnarray}  \label{beta-total}
\beta^{total}_k = \beta_k +
 e^{-1}\gamma^3 ({\bf a}^{-1})^i_{\ k}
Q_i\ ,
\end{eqnarray}
where
\begin{eqnarray} \label{genforce}
Q_i = - \frac{\partial U}{\partial q^i} +
\frac{d}{dt}\frac{\partial U}{\partial \dot{q}^i}
\end{eqnarray}
is a vector of the generalized force in the laboratory
frame.
The complete Hamiltonian {\bf H} on the constraint manifold
reads:
\begin{eqnarray}     \label{H-external}
{\bf H} = p_kv^k + \gamma^{-1}{\cal H} + U =
\gamma \left({\cal H} + v^k{\cal
P}_k\right) + U - v^k\frac{\partial U}{\partial \dot{q}^k}\ .
\end{eqnarray}
Obviously, the constraint manifold $\overline{{\cal P}}$ is
endowed with a weak symplectic form $\overline{\Omega}$ and
one can prove the following
\begin{TH}   \label{TH3}
The Hamiltonian vector field $X_{{\bf H}}$ corresponding to
(\ref{H-external}) is well defined on the  submanifold
${\cal P}^*$ defined by the following non-homogeneous
boundary condition
\end{TH}
\begin{eqnarray}   \label{extfund}
\mbox{DP}(4\pi m\Psi^2 + e^2\Psi^2,_3)(0) =
 - e\gamma^3 ({\bf a}^{-1})^k_{\ i} Q_k\frac{x^i}{r}\ .\nonumber
\end{eqnarray}
The proof is analogous to the proof of Theorem~\ref{TH1}.

Finally, the Hamiltonian structure for the particle in an
external potential is a triple $({\cal P}^*,\Omega^*,{\bf
H})$, where $\Omega^*$ is a reduction of
$\overline{\Omega}$ to ${\cal P}^*$. Due to the
Theorem~\ref{TH3} this structure is well defined, i.e.  the
initial data $(\Psi,\chi;{\bf q},{\bf v})$ for the
radiation field and for the particle uniquely determine the
entire history of the system if the external potential is
given.

As an example consider the particle interacting with an
external electromagnetic field $f^{ext}_{\mu\nu}$. The
generalized potential is given by:
\begin{eqnarray}
U({\bf q},\dot{\bf q},t) = eA^{ext}_0({\bf q},t) -
e\dot{\bf q}{\bf A}^{ext}({\bf q},t) \ ,\nonumber
\end{eqnarray}
where $A^{ext}_0$ and ${\bf A}^{ext}$ stand for the
four-potential of the external field in the laboratory
frame. The generalized force (\ref{genforce}) in terms of
the laboratory-frame components ${\cal E}_i$ and ${\cal
B}_i$ of the external field now reads:
\begin{eqnarray}
Q_i = e\left( {\cal E}_i({\bf q},t) +
\epsilon_{ijk}v^j{\cal B}^k({\bf q},t) \right)\ .\nonumber
\end{eqnarray}
In this case
\begin{eqnarray}
U - v^k\frac{\partial U}{\partial \dot{q}^k} = eA^{ext}_0
\nonumber
\end{eqnarray}
and the complete Hamiltonian reads
\begin{eqnarray}     \label{H-external1}
{\bf H}({\bf q},{\bf v};\mbox{fields}) =
\gamma \left({\cal H} + v^k{\cal P}_k\right) +
eA^{ext}_0\ .
\nonumber
\end{eqnarray}
Moreover, since $U$ is linear in the velocity, we may
easily express {\bf v} in terms of {\bf q}, {\bf p} and
fields (radiation and external). One obtains formula
analogous to (\ref{vk}) with $p_k$ replaced by $p_k +
eA_k^{ext}$ which leads to the following expression for {\bf
H}:
\begin{eqnarray}
{\bf H}({\bf q},{\bf p};\mbox{fields}) = \sqrt{{\cal H}^2 +
({\bf p} + e{\bf A}^{ext})^2 -      \mbox{\boldmath   $\cal
P$}^2}   + eA_0^{ext}\ .
\nonumber
\end{eqnarray}

Finally, let us observe that the ``commutation relations''
between particle's variables {\bf q}, {\bf v} and field
variables $(\Psi,\chi)$ have the same form as
in Appendix~\ref{Commutation-relations} with $\beta$ replaced by
$\beta^{total}$ defined in (\ref{beta-total}).

\section*{Appendixes}
\addcontentsline{toc}{section}{\protect\numberline{}Appendixes}
\appendix
\def\theequation{\thesection.\arabic{equation}}

\section{Canonical formalism for a 2-nd order
Lagrangian theory}
\setcounter{equation}{0}
\label{CanStruc}

Consider a theory described by the 2-nd order lagrangian $L
= L(q,\dot{q},\ddot{q})$ (to simplify the notation we skip
the index ``$i$'' corresponding to different degrees of
freedom $q^i$; extension of this approach to higher order
Lagrangians is straightforward).  Introducing auxiliary
variables $v = \dot{q}$ we can treat our theory as a 1-st
order one with lagrangian constraints $\phi := \dot{q} - v
= 0$ on the space of lagrangian variables
$(q,\dot{q},v,\dot{v})$.  Dynamics is generated by the
following relation:
\begin{eqnarray}   \label{rel1}
d\, L(q,v,\dot{v}) =
\frac{d}{dt}\big(p\,dq + \pi \,dv\big) = \dot{p}\,dq +
p\,d\dot{q} +
\dot{\pi}\,dv + \pi \, d{\dot v} \ .
\end{eqnarray}
where $(p, \pi)$ are momenta canonically conjugate to $q$
and $v$ respectively.  Because $L$ is defined only on the
constraint submanifold, its derivative $dL$ is not uniquely
defined and has to be understood as a collection of {\em
all the covectors} which are compatible with the derivative
of the function along constraints. This means that the left
hand side is defined up to $ \mu (\dot{q} - v)$, where
$\mu$  are Lagrange multipliers corresponding to
constraints $\phi = 0$ . We conclude that $p = \lambda$ is
an arbitrary covector and (\ref{rel1}) is equivalent to the
system of dynamical equations:
\begin{eqnarray*}
 \pi = \frac{\partial L}{\partial \dot{v}} \, ,\ \ \ \ \ \
\dot{p} = \frac{\partial L}{\partial q}\, ,\ \ \ \ \ \
\dot{\pi} =
\frac{\partial L}{\partial v} - p \, .
\end{eqnarray*}
The last equation implies the definition of the canonical
momentum $p$:
\begin{eqnarray}   \label{p}
p = \frac{\partial L} {\partial v} - \dot{\pi} =
\frac{\partial L} {\partial v} - \frac{d}{dt}\left(
\frac{\partial L}{\partial
\dot{v}} \right)\ .
\end{eqnarray}
We conclude, that equation
\begin{eqnarray}
\dot{p} = \frac{d}{dt}\left(\frac{\partial L}{\partial v}\right) -
\frac{d^{2}}{dt^{2}}\left(\frac{\partial L}{\partial \dot{v}}
\right) \nonumber \end{eqnarray}
is equivalent, indeed, to the  Euler-Lagrange equation:
\begin{eqnarray} \label{EL}
\frac{\delta L}{\delta q} := \frac{d^{2}}{dt^{2}}\left(\frac
{\partial L}{\partial \dot{v}}\right) -
\frac{d}{dt}\left(\frac {\partial L}{\partial v}\right) +
\frac{\partial L}{\partial q} = 0\, .
\end{eqnarray}
The Hamiltonian description  is
obtained from the Legendre transformation applied to
(\ref{rel1}):
\begin{eqnarray}      \label{rel2}
-dH = \dot{p}\,dq - \dot{q}\,dp +\dot{\pi}\,dv
-\dot{v}\,d\pi\, ,
\end{eqnarray}
where $H(q,p,v,\pi) = p\,v +\pi \,\dot{v} - L(q,v, \dot{v}
)$. In this formula we have to insert $\dot{v} = \dot{v}
(q,v,\pi )$, calculated from  equation $\pi =
\frac{\partial L}{\partial
\dot{v}}$. Let us observe that $H$ is linear with respect to the
momentum $p$. This is a characteristic feature of the 2-nd
order theory.

Euler-Lagrange equations (\ref{EL}) are of  4-th order. The
corresponding 4 Hamiltonian equations have, therefore, to
describe the evolution of $q$ and its derivatives up to
third order. Due to Hamiltonian equations implied by
relation (\ref{rel2}), the information about succesive
derivatives of $q$ is carried by $(v,\pi ,p)$:

\begin{itemize}

\item v describes $\dot{q}$  \begin{eqnarray}
\dot{q} = \frac{\partial H}{\partial p} \equiv v
\nonumber\end{eqnarray}
hence, the constraint $\phi = 0$ is reproduced due to
linearity of $H$ with respect to $p$,

\item $\pi$ contains information about $\ddot{q}$:
\begin{eqnarray}
\dot{v} = \frac{\partial H}{\partial \pi}\, ,
\nonumber\end{eqnarray}

\item   $p$ contains information about $\stackrel{...}{q}$
\begin{eqnarray}
\dot{\pi} = - \frac{\partial H}{\partial v} = \frac{\partial
L}{\partial v} - p \, ,\nonumber\end{eqnarray}

\item  the true dynamical equation equals
\begin{eqnarray}
\dot{p} = - \frac{\partial H}{\partial q} = \frac{\partial L}
{\partial q} \, . \nonumber\end{eqnarray}
\end{itemize}


\section{Proof of Theorem 4}
\setcounter{equation}{0}
\label{PROOF}

Observe, that due to (\ref{expansion})
 the field $\Psi^2$ has the following expansion in the vicinity of the
particle:
\begin{eqnarray}
\Psi^2(r) = \frac{\alpha}{4\pi}  + r\beta + O(r^2)\ ,
\end{eqnarray}
where $O(r^2)$ denotes terms vanishing for $r\rightarrow 0$
like $r^2$ or faster. We shall prove that the Hamiltonian
vector field $X_{{\bf H}}$ corresponding to the Hamiltonian
(\ref{H-qv}) is well defined if and only if the
fundamental equation (\ref{fund}) is satisfied.  The field $X_{{\bf H}}$
is defined by
\begin{eqnarray}        \label{XH-Omega}
X_{{\bf H}} \con \Omega^* = - \delta{\bf H}\ ,
\end{eqnarray}
where $\Omega^*$ is given by (\ref{Omega*}) and $X_{{\bf
H}}$ may be written in terms of coordinates as follows:
\begin{eqnarray}   \label{XH}
X_{{\bf H}} = X^i\frac{\partial}{\partial q^i} +
Y^i\frac{\partial}{\partial v^i} + W^A\frac{\delta}{\delta
\Psi^A} + Z_A\frac{\delta}{\delta \chi_A}\ .
\end{eqnarray}
 Using (\ref{XH-Omega}) and (\ref{XH}) we get
\begin{eqnarray}
\label{H-q}
-\frac{\partial{\bf H}}{\partial q^k} &=& Y^l\left[
\frac{\partial(\gamma v_k)}{\partial v^l}{\cal H} +
\frac{\partial {\bf a}^n_{\ k}}{\partial v^l}{\cal P}_n\right] +
W^A\left[\gamma v_k \frac{\delta{\cal H}}{\delta\Psi^A} +
{\bf a}^n_{\ k}\frac{\delta{\cal P}_n}{\delta\Psi^A}\right]
\nonumber\\ &+& Z_A\left[\gamma v_k \frac{\delta{\cal
H}}{\delta\chi_A} + {\bf a}^n_{\ k}\frac{\delta{\cal
P}_n}{\delta\chi_A}\right] \ , \\
\label{H-v}
-\frac{\partial{\bf H}}{\partial v^l} &=& -X^k\left[
\frac{\partial(\gamma v_k)}{\partial v^l}{\cal H} +
\frac{\partial {\bf a}^n_{\ k}}{\partial v^l}{\cal P}_n\right]
\nonumber\\  &+&
Y^k\left[ \left(\frac{\partial(\gamma {\bf a}^n_{\ k})}{\partial
v^l} -
\frac{\partial(\gamma {\bf a}^n_{\ l})}{\partial v^k}\right){\cal R}_n +
\left(\frac{\partial \om_{km}}{\partial v^l} -
\frac{\partial \om_{lm}}{\partial v^k}\right)S^m\right]
\nonumber\\ &-& W^A\left[\gamma {\bf a}^n_{\ l} \frac{\delta{\cal
R}_n}{\delta\Psi^A} +
\om_{lm} \frac{\delta S^m}{\delta\Psi^A}\right]
- Z_A\left[\gamma {\bf a}^n_{\ l} \frac{\delta{\cal
R}_n}{\delta\chi_A} +
\om_{lm} \frac{\delta S^m}{\delta\chi_A}\right]\ , \\
\label{H-Psi}
-\frac{\delta{\bf H}}{\delta\Psi^A} &=& -X^k \left[\gamma
v_k
\frac{\delta{\cal H}}{\delta\Psi^A} + {\bf a}^n_{\
k}\frac{\delta{\cal P}_n}{\delta\Psi^A}\right] +
Y^k\left[\gamma {\bf a}^n_{\ k}
\frac{\delta{\cal R}_n}{\delta\Psi^A} +
\om_{km} \frac{\delta S^m}{\delta\Psi^A}\right] + Z_A\ ,\\
\label{H-chi}
-\frac{\delta{\bf H}}{\delta\chi_A} &=& -X^k \left[\gamma
v_k
\frac{\delta{\cal H}}{\delta\chi_A} + {\bf a}^n_{\
k}\frac{\delta{\cal P}_n}{\delta\chi_A}\right] +
Y^k\left[\gamma {\bf a}^n_{\ k}
\frac{\delta{\cal R}_n}{\delta\chi_A} +
\om_{km} \frac{\delta S^m}{\delta\chi_A}\right] - W^A\ .
\end{eqnarray}
Using the definition of {\bf H} (see (\ref{H-qv})) we have
\begin{eqnarray}
\frac{\partial{\bf H}}{\partial q^k} &=& 0\ ,\\
\frac{\partial{\bf H}}{\partial v^l} &=&
\gamma^3v_k({\cal H} + v^l{\cal P}_l) + \gamma{\cal P}_k\ ,\\
\frac{\delta{\bf H}}{\delta\Psi^A} &=&
\gamma \left[ \frac{\delta{\cal H}}{\delta\Psi^A} +
v^l\frac{\delta{\cal P}_l}{\delta\Psi^A}\right] \ ,\\
\frac{\delta{\bf H}}{\delta\chi_A} &=&
\gamma \left[ \frac{\delta{\cal H}}{\delta\chi_A} +
v^l\frac{\delta{\cal P}_l}{\delta\chi_A}\right] \ .
\end{eqnarray}
Now, calculating $W^A$ and $Z_A$ from (\ref{H-chi}) and
(\ref{H-Psi}) respectively, inserting them to
(\ref{H-v}) and taking into account the Poincar\`e algebra relations, we get
\begin{eqnarray}
\lefteqn{ - A_l{\cal H} + B_l^{\ n}{\cal P}_n -
\gamma^2 {\bf a}^n_{\ l}v^kg_{nk}\left(m-\frac{e^2}{6\pi r_0}\right)
}\nonumber\\ &=& X^k\left[  C_{lk}{\cal H} + D_{lk}^{\ n}{\cal
P}_n    -
\gamma {\bf a}^n_{\ l}{\bf a}^r_{\ k}g_{nr}
\left(m-\frac{e^2}{6\pi r_0}\right) \right]
+ Y^k\left[ E_{lk}^{\ \ n}{\cal R}_n + F_{lkm}S^m \right]\ ,
\end{eqnarray}
where the following 3-dimensional objects depending  upon
the velocity {\bf v}  are introduced:
\begin{eqnarray*}
A_l &=& \gamma^3v_l - \gamma^2 {\bf a}^n_{\ l}v^kg_{nk}\ ,\\
B_l^{\ n} &=& - \gamma^3v_lv^n - \gamma\delta_l^{\ n} +
 \gamma^2 {\bf a}^n_{\ l}
- \om_{lm}v^k\epsilon^{nm}_{\ \ k}\ ,\\
C_{lk} &=& -\gamma(g_{kl}+\gamma^2v_kv_l) + \gamma {\bf a}^n_{\ l}{\bf a}^r_{\
k}g_{nr}\ ,\\
\label{Dln}
D_{lk}^{\ n} &=& - \frac{\partial {\bf a}^n_{\ k}}{\partial v^l}
+\gamma^2 v_k{\bf a}^n_{\ l}
- \om_{lm}{\bf a}^i_{\ k}\epsilon_i^{\ mn}\ ,\\
E_{lk}^{\ \ n} &=&
\left(\frac{\partial(\gamma {\bf a}^n_{\ k})}{\partial v^l} -
\frac{\partial(\gamma {\bf a}^n_{\ l})}{\partial v^k}\right)
+ \gamma({\bf a}^i_{\ l}\om_{km} - {\bf a}^i_{\ k}\om_{lm})\epsilon_i^{\ mn}\ , \\
F_{lkm} &=&
\left(\frac{\partial \om_{km}}{\partial v^l} - \frac{\partial
\om_{lm}}{\partial v^k}\right)
+ \om_{lr}\om_{kj}\epsilon^{rj}_{\ \ m}\ .
\end{eqnarray*}
Using the following properties of the function $\varphi(\tau)$:
\begin{eqnarray*}
2\varphi(\tau) - (1-\tau)^{-1} + \tau\varphi^2(\tau) &=& 0\ ,\\
2\varphi'(\tau) - (1-\tau)^{-1}\varphi(\tau) - \varphi^2(\tau) &=& 0\ ,
\end{eqnarray*}
and the identity
\begin{eqnarray}
v^i(\epsilon_{ikl}v_m + \epsilon_{ilm}v_k +  \epsilon_{imk}v_l)
= {\bf v}^2\epsilon_{klm}\ ,\nonumber
\end{eqnarray}
one easily shows that
\begin{eqnarray}
A_l =0\ , \ \ B_l^{\ n}=0\ ,\ \ C_{lk}=0\ ,\ \ D_{lk}^{\ n}=0\ ,\ \
E_{lk}^{\ \ n}=0\ ,\ \ F_{lkm} =0\ .\nonumber
\end{eqnarray}
Now, taking into account that
\begin{eqnarray}
{\bf a}^n_{\ l}v_n &=& \gamma v_l\ ,\nonumber\\
\label{bb}
{\bf a}^n_{\ l}{\bf a}^r_{\ k}g_{nr} &=& g_{lk} + \gamma^2v_lv_k \ ,
\nonumber
\end{eqnarray}
we obtain finally the following equation for $X^k$:
\begin{eqnarray}
X^k(g_{kl}+\gamma^2v_kv_l) = \gamma^2v_l\ .\nonumber
\end{eqnarray}
The matrix $(g_{kl}+\gamma^2v_kv_l)$ is nonsingular and its inverse equals to
$(g^{kl} - v^kv^l)$. Therefore
\begin{eqnarray}
X^k = \gamma^2(g^{kl} - v^kv^l)v_l = v^k\ .\nonumber
\end{eqnarray}

Now, let us consider equation (\ref{H-q}) and apply the same strategy as in
the case of equation (\ref{H-v}). Calculate $W^A$ and $Z_A$ from (\ref{H-chi})
and (\ref{H-Psi}) respectively and insert them into (\ref{H-q}).
 Then, inserting $X^l=v^l$ and using once more the Poincar\`e algebra structure
  one gets:
\begin{eqnarray}
\lefteqn{
\gamma(v_kv^l{\bf a}^n_{\ l} - v^2{\bf a}^n_{\ k} + {\bf a}^n_{\ k} -
 \gamma v^nv_k)
\left(\frac{e}{6\pi r_0}\alpha_n + e\beta_n\right) } \nonumber\\ &=&
 Y^l \left[ \gamma(g_{kl} + \gamma^2v_kv_l){\cal H}
- {\bf a}^n_{\ l}{\bf a}^r_{\ k}g_{nr}\left({\cal H}-m +\frac{e^2}{6\pi
r_0}\right) \right]\ ,
\end{eqnarray}
Now, observing that
\begin{eqnarray}
v_kv^l{\bf a}^n_{\ l} - v^2{\bf a}^n_{\ k} + {\bf a}^n_{\ k} - \gamma v^nv_k
= \gamma^{-2}{\bf a}^n_{\ k}  \nonumber
\end{eqnarray}
 we finally obtain equation for $Y^l$:
\begin{eqnarray}  \label{Yl}
Y^l(g_{kl} + \gamma^2v_kv_l)
\left(\frac{e^2}{6\pi r_0} - m\right) =
\gamma^{-2}{\bf a}^n_{\ k}\left(\frac{e}{6\pi r_0}\alpha_n + e\beta_n\right)\ .
\end{eqnarray}
The Hamiltonian vector field $X_{{\bf H}}$ is well defined if and only if
the singular terms proportional to $r_0^{-1}$ cansel out, i.e.
\begin{eqnarray}
eY^l(g_{kl} + \gamma^2v_kv_l)
= \gamma^{-2}{\bf a}^n_{\ k}\alpha_n \ ,\nonumber
\end{eqnarray}
and therefore
\begin{eqnarray}
Y^l = \frac 1e \gamma^{-2}(g^{lk} - v^lv^k){\bf a}^n_{\ k}\alpha_n
= \frac 1e\gamma^{-2} (g^{ln} -
\gamma^{-1}\varphi\,v^lv^n)\alpha_n\ .\nonumber
\end{eqnarray}
When the above equation holds then from (\ref{Yl})
\begin{eqnarray}
-mY^l(g_{kl} + \gamma^2v_kv_l)
= e \gamma^{-2}{\bf a}^n_{\ k}\beta_n \ ,\nonumber
\end{eqnarray}
and finally
\begin{eqnarray}
m\alpha_k = -e^2\beta_k\ ,\nonumber
\end{eqnarray}
which ends the proof of Theorem~\ref{TH1}.

\section{Commutation relations}
\label{Commutation-relations}
\setcounter{equation}{0}

The complete set of ``commutation relations'' read:
\begin{eqnarray*}   \label{q-q}
       \{ q^{k},q^{l}\}^{*} & = & 0\ , \\ \{
q^{k},v^{l}\}^{*} & = & 0\ , \\ \{ v^{k},v^{l}\}^{*} & = &
0\ , \\ \{ q^{k},\Psi^{1}({\bf x})\}^{*} & = & 0\ , \\ \{
q^{k},\chi_{1}({\bf x})\}^{*} & = & 0\ , \\ \{
q^{k},\Psi^{2}({\bf x})\}^{*} & = & 0\ , \\ \{
q^{k},\chi_{2}({\bf x})\}^{*} & = &
\frac{3}{2e}\gamma^2 ({\bf a}^{-1})^k_{\ l}
 \frac{x^l}{|{\bf x}|}\delta(|{\bf
x}|)\ ,\\ \{ v^{k},\Psi^{1}({\bf x})\}^{*} & = & 0\ , \\ \{
v^{k},\chi_{1}({\bf x})\}^{*} & = & 0\ , \\ \{
v^{k},\Psi^{2}({\bf x})\}^{*} & = & - \gamma ({\bf a}^{-1})^k_{\ l}
 \frac{x^l}{|{\bf
x}|}\delta(|{\bf x}|)\ , \\ \{ v^{k},\chi_{2}({\bf
x})\}^{*} & = & - \frac{3}{2m}\gamma^{-3} v^{[k}\beta^{j]}
 {\bf a}^l_{\ j}
\frac{x_l}{|{\bf
x}|}\delta(|{\bf x}|)\ ,
\end{eqnarray*}
and for field functionals:
\begin{eqnarray*}
     \{\Psi^{1}({\bf x}),\Psi^{1}({\bf y})\}^{*} & = & 0\ , \\
 \{\Psi^{1}({\bf x}),\chi_{1}({\bf y})\}^{*} & = &
\delta^3({\bf x} - {\bf y})\ , \\
\{\Psi^{1}({\bf
x}),\Psi^{2}({\bf y})\}^{*} & = & - \frac{3}{e}\frac{\delta
{\cal K}_{l}}{\delta \chi_{1}({\bf x})}\ \frac{y^l}{|{\bf
y}|}\delta(|{\bf y}|)\ , \\
\{\Psi^{1}({\bf
x}),\chi_{2}({\bf y})\}^{*} & = & -
\frac{3}{2e}\frac{\delta
\Lambda_{l}}{\delta \chi_{1}({\bf x})}\
\frac{y^l}{|{\bf y}|}\delta(|{\bf y}|)\ , \\
\{\chi_{1}({\bf x}),\chi_{1}({\bf y})\}^{*} & = & 0\ , \\
\{\chi_{1}({\bf x}),\Psi^{2}({\bf y})\}^{*} & = &
\frac{3}{e}\frac{\delta {\cal K}_{l}}{\delta \Psi^{1}({\bf
x})}\
\frac{y^l}{|{\bf y}|}\delta(|{\bf y}|)\ , \\
\{\chi_{1}({\bf x}),\chi_{2}({\bf y})\}^{*} & = &
\frac{3}{2e}\frac{\delta
\Lambda_{l}}{\delta \Psi^{1}({\bf x})}\
\frac{y^l}{|{\bf y}|}\delta(|{\bf y}|)\ , \\
\{\Psi^{2}({\bf x}),\Psi^{2}({\bf y})\}^{*} & = &
\frac{3}{e}\left(
\frac{\delta {\cal K}_{l}}{\delta
\chi_{2}({\bf x})}\
\frac{y^l}{|{\bf y}|}\delta(|{\bf y}|) -
\frac{\delta {\cal K}_{l}}{\delta \chi_{2}({\bf y})}\
\frac{x^l}{|{\bf x}|}\delta(|{\bf x}|) \right) \ ,\\
\{\chi_{2}({\bf x}),\chi_{2}({\bf y})\}^{*} & = &
\frac{3}{2e}\left(
\frac{\delta \Lambda^{l}}{\delta \chi_{2}({\bf x})}\
\frac{y^l}{|{\bf y}|}\delta(|{\bf y}|) -
\frac{\delta \Lambda^{l}}{\delta \chi_{2}({\bf y})}\
\frac{x^l}{|{\bf x}|}\delta(|{\bf x}|) \right) \ ,\\
\label{Psi2-chi2}
\{\Psi^{2}({\bf x}),\chi_{2}({\bf y})\}^{*} & = &
\delta^3({\bf x}-{\bf y}) - \frac{9m}{2e^2}\frac{x^k}{|{\bf
x}|}
\frac{y_k}{|{\bf y}|}\delta(|{\bf x}|) \delta(|{\bf
y}|)  \nonumber\\ &-& \frac{3}{2e}
\frac{\delta \Lambda^{l}}{\delta \chi_{2}({\bf x})}\
\frac{y^l}{|{\bf y}|}\delta(|{\bf y}|) -
\frac{3}{e}
\frac{\delta {\cal K}_{l}}{\delta
\Psi^{2}({\bf y})}\
\frac{x^l}{|{\bf x}|}\delta(|{\bf x}|)\ .
\end{eqnarray*}
Functionals ${\cal K}_l$ and $\Lambda_l$ are defined as
follows:
\begin{eqnarray*}   \label{defK}
{\cal K}_{i} & := & {\cal R}_{i} + \gamma^{-1}
\varphi\,v^{l}\,\epsilon_{lim}S^{m}\ -
\frac{e}{4\pi} \int_{\Sigma}\lambda \frac{x_i}{r^3}\Psi^2\ ,\\
\label{defL}
\Lambda_{i} & := & {\cal P}_{i} + v_{i}\left( {\cal H} +
\frac{e}{m}
{\cal K}^{l}\beta_{l} \right)- \left(mv_i - \frac{e}{2\pi}
\int_{\Sigma} \lambda \frac{x_i}{r^3} \chi_2 \right) \ .
\end{eqnarray*}

\section{Derivation of the formula ${\bf v}={\bf v}({\bf p})$.}
\label{velocity-momentum}
\setcounter{equation}{0}

From (\ref{pk}) we have
\begin{eqnarray}   \label{ck}
c_k := p_k - {\cal P}_k = \left( \gamma {\cal
H} +
\varphi\,v^l{\cal P}_l \right)v_k\ .
\end{eqnarray}
Now, let $x=v^2$, $A={\cal H}\sqrt{c_kc^k}={\cal H}|{\bf
c}|$, $B=c_l{\cal P}^l$ and $C=c_lp^l$. From (\ref{ck})
vectors ${\bf c} = (c_k)$ and {\bf v} are parallel and
therefore
\begin{eqnarray}  \label{parallel}
v_k = \frac{|{\bf v}|}{|{\bf c}|}c_k\ .
\end{eqnarray}
Multiplying both sides of (\ref{pk}) by $v^k$ we obtain
\begin{eqnarray}            \label{pkvk}
p_kv^k = \gamma ({\cal H} + {\cal P}_kv^k) \ .
\end{eqnarray}
Moreover,
\begin{eqnarray}   \label{pkvk-1}
{\cal P}_kv^k = \frac{|{\bf v}|}{|{\bf c}|}c^k{\cal P}_k =
\frac{|{\bf v}|}{|{\bf c}|}B\ .
\end{eqnarray}
and
\begin{eqnarray}   \label{pkvk-2}
p_kv^k = \frac{|{\bf v}|}{|{\bf c}|}c^kp_k = \frac{|{\bf
v}|}{|{\bf c}|}C\ .
\end{eqnarray}
Therefore, inserting (\ref{pkvk-1}) and (\ref{pkvk-2}) into
(\ref{pkvk}) we obtain
\begin{eqnarray}
C = A\sqrt{\frac{x}{1-x}} + B\frac{1}{\sqrt{1-x}}\ .
\end{eqnarray}
The above equation is a square equation for $\sqrt{x}$:
\begin{eqnarray}      \label{square-eq}
(A^2+C^2)x + 2AB\sqrt{x} + (B^2-C^2) = 0\ .
\end{eqnarray}
The discriminant $\Delta$ for (\ref{square-eq}) equals:
\begin{eqnarray}
\Delta = 4C^2(A^2+B^2-C^2)\ .\nonumber
\end{eqnarray}
 Calculating $x$ and
inserting into (\ref{parallel}) we finally obtain
(\ref{vk}).

\section*{Acknowledgements}

I am very much indebted to Prof. J. Kijowski from the
Center of Theoretical Physics Polish Academy of Sciences (Warsaw) for many
highly stimulating discussions about the problem of interacting particles
and fields. 

Many thanks are due to Prof. A. Kossakowski from the N. Copernicus
University for very interesting discussions about self-interacting problem
and to  Prof. H. R\"omer from the Freiburg University for his
interests in this work.

Finally, I thank the Alexander von Humboldt Foundation for the financial
 support.

\end{sloppypar}

\end{document}